\documentclass[%
 reprint,
superscriptaddress,
nofootinbib,
 amsmath,amssymb,
 aps,
 prl,
]{revtex4-2}
\usepackage{layouts}
\usepackage{graphicx, subfigure}% Include figure files
\usepackage{dcolumn}% Align table columns on decimal point
\usepackage{bm}% bold math
\usepackage{pxtx-cal}% cal math
\usepackage{mathtools}
\usepackage{amsmath}
\usepackage{amssymb}
\usepackage{pgfplots}
\usepackage{float}
\usepgfplotslibrary{fillbetween}
\pgfplotsset{compat=1.16}
\usetikzlibrary{fit,backgrounds,positioning,shapes,shapes.callouts}
    \usepackage[scaled=0.83]{helvet}
    \usepackage{marvosym,pifont}
    \usepackage[T1]{fontenc}
    \usepackage[utf8]{inputenc}
\usepackage{hyperref}% add hypertext capabilities

\begin{document}

\title{Coherent interaction-free detection of microwave pulses  with a superconducting circuit}

\author{Shruti Dogra}
 \email[]{shruti.dogra@aalto.fi}

\affiliation{QTF  Centre  of  Excellence, 
 Department of Applied Physics, Aalto University, FI-00076 Aalto, Finland\\}

\author{John J. McCord}
 \email[]{john.mccord@aalto.fi}

\affiliation{QTF  Centre  of  Excellence, 
	Department of Applied Physics, Aalto University, FI-00076 Aalto, Finland\\}

\author{Gheorghe Sorin Paraoanu}
 \email[]{sorin.paraoanu@aalto.fi}

\affiliation{QTF  Centre  of  Excellence, 
	Department of Applied Physics, Aalto University, FI-00076 Aalto, Finland\\}

\date{\today}% 

\begin{abstract}
The interaction-free measurement is a fundamental quantum effect whereby the presence of a photosensitive object is determined without irreversible photon absorption. Here we propose the concept of coherent interaction-free detection and demonstrate it experimentally using a three-level superconducting transmon circuit. In contrast to standard interaction-free measurement setups, where the dynamics involves a series of projection operations, our protocol employs a fully coherent evolution that results, surprisingly, in a higher probability of success. We show that it is possible to ascertain the presence of a microwave pulse resonant with the second transition of the transmon, while at the same time avoid exciting the device onto the third level. Experimentally, this is done by using a series of Ramsey microwave pulses coupled into the first transition and monitoring the ground-state population.
\end{abstract}

\pacs{Valid PACS appear here}

\maketitle

\section*{Introduction}

Since the inception of quantum mechanics, the quest to understand measurements has been a rich source of intellectual fascination. In 1932 von Neumann provided the paradigmatic projective  model \cite{von-Neumann}
	while in recent times a lot of research has been done on alternative forms and generalizations such as partial measurements and their reversal \cite{Katz_2006, Katz_2008, Paraoanu_2011, Paraoanu_2011_2}, 
	weak measurements  \cite{Aharonov_1988, Aharonov_2014, Dressel_2014, Hatridge_2013} and
	their complex weak values \cite{ Groen_2013, Campagne-Ibarcq_2014}, observation of quantum trajectories \cite{Murch_2013, Roch_2014},
	and simultaneous measurements of non-commuting observables \cite{Arthurs_1965, Hacohen-Gourgy2016, Piacentini_2016}.

The interaction-free measurements belong to the class of quantum hypothesis testing, where the existence of an event (for example the presence of a target in a region of space) is assessed. In a nutshell, the interaction-free detection protocol \cite{Elitzur_1993} provides a striking illustration of the concept of negative-results measurements of Renninger \cite{Renninger} and Dicke \cite{Dicke}. The very presence of an ultrasensitive object in one of the arms of a Mach-Zehnder interferometer modifies the output probabilities even when no photon has been absorbed by the object. The detection efficiency can be enhanced by using the quantum Zeno effect \cite{peres-ajp-1980} through repeated no-absorption ``interrogations'' of the object \cite{Kwiat_1995, Kwiat_1999, Ma2014, Peise2015} -- a protocol which we will refer to as ``projective''.
Other detection schemes in the hypothesis testing class have been advanced, most notably quantum illumination 
\cite{Lloyd_2008, Tan_2008},
ghost imaging -- where the imaging photons have not interacted with the imaged object \cite{Klyshko_2007, Pittman_1995, Strekalov_1995}, and imaging with undetected photons \cite{Lemos_2014, Lahiri_2015}. The interaction-free concept has touched off a flurry of research in the foundations of quantum mechanics, for example
	the Hardy paradox \cite{Hardy_1992}, non-local effects between distant atoms exchanging photons \cite{Aharonov_2018}, and quantum engines \cite{Elouard_2020}.

Here we describe and demonstrate experimentally a hypothesis-testing protocol that employs repeated coherent interrogations instead of projective ones. In this protocol, the task is to detect the presence of a microwave pulse in a transmission line using a resonantly-activated detector realized as a transmon three-level device. We require that at the end of the protocol the detector has not irreversibly absorbed the pulse, as witnessed by a non-zero occupation of the second excited state. Clearly this task cannot be achieved with a classical absorption-based detector (e.g., a bolometer) or by using a simple two-level system as a detector. Our protocol is fundamentally different from the quantum Zeno interaction-free measurement: while in the latter case the mechanism of detection is the suppression of the coherent evolution by projection on the interferometer path that does not contain the object, in our protocol the evolution of the state of the superconducting circuit remains fully coherent.  Surprisingly, this coherent addition of amplitude probabilities results in a higher probability of successful detection.
	
This concept can be implemented in other experimental platforms where a three-level system is available. We note that projective interaction-free measurements have found already applications in optical imaging \cite{White1998}, counterfactual communication \cite{salih-prl-2013,Vaidman_2015, Cao2017,vaidman-pra-2019, Walther2019, Aharonov2021}, ghost-imaging \cite{Zhang_2019, hans-npj-2021}, detection of noise in mesoscopic physics \cite{Chirolli_2010}, cryptographic key distribution \cite{Noh2009,zheng-pra-2020}, and measurement-driven engines \cite{andrew-found2020}. We expect that our coherent version will be similarly adapted to these nascent fields.

%\section{Experimental implementation}

\begin{figure}[ht]
\centering
\includegraphics[width=0.8\linewidth]{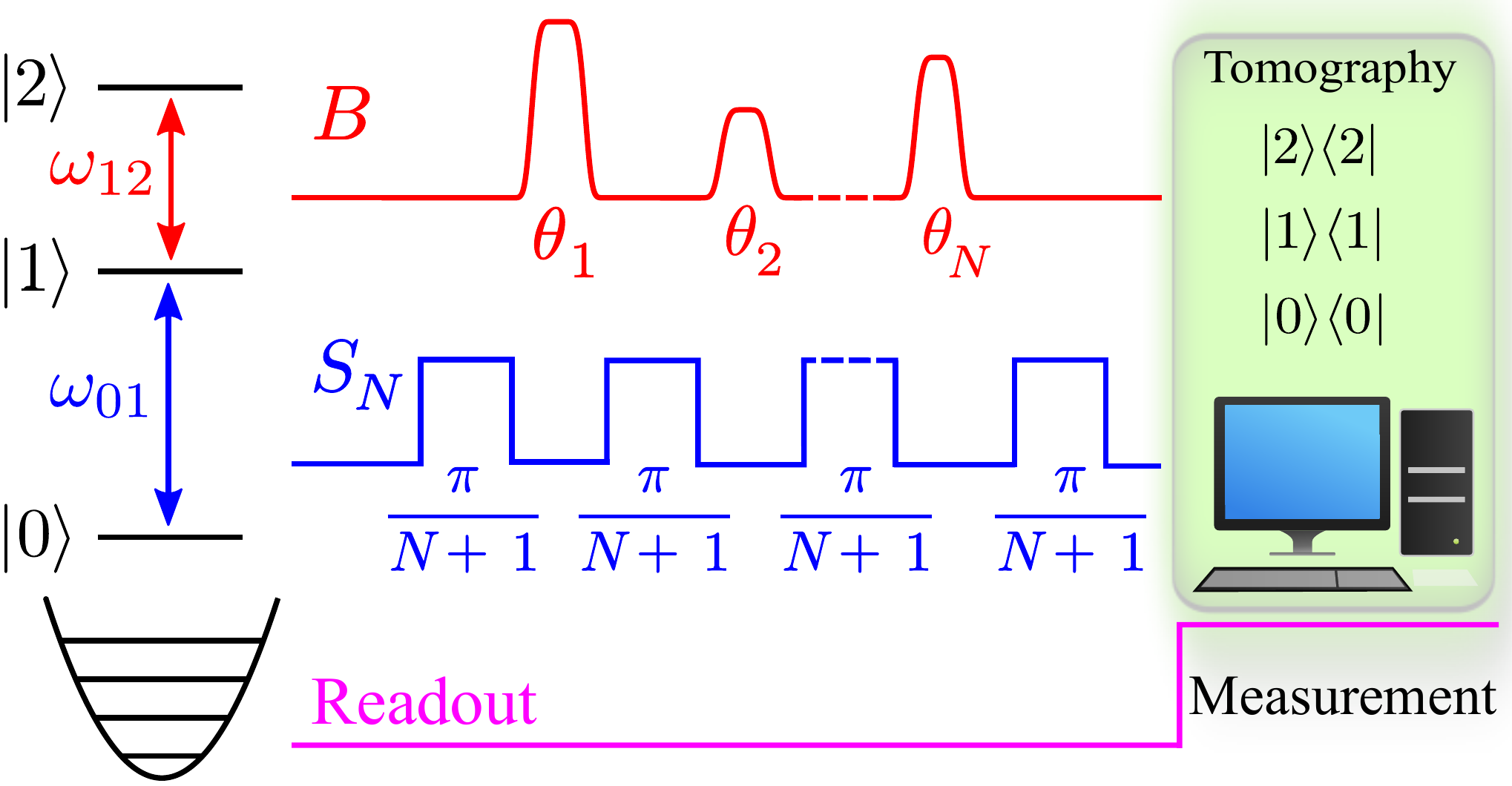}
\caption{{\bf Coherent interaction-free detection.} Schematic of the protocol, where $S_{N}$ and $B$ microwave pulses are shown in blue and red, respectively, along with the probe pulse for readout. 
}
\label{fig-pulses}
\end{figure}

In our experiments, we realize a series of $N$ Ramsey-like sequences by applying beam-splitter unitaries $S_{N}$ to the lowest two energy levels of a superconducting transmon. This creates the analog of the standard Mach-Zehnder spatial setup
in a time-domain configuration \cite{Paraoanu_2006}.
The microwave pulses of strength $\theta_j$ that we wish to detect -- which we will refer to as $B$-pulses --  couple resonantly into the next higher transition, see Fig.~\ref{fig-pulses}. Specifically, let us denote the first three levels of the transmon by $|0\rangle$, $|1\rangle$, and $|2\rangle$ and the asymmetric Gell-Mann generators of SU(3) by $\sigma^{y}_{kl} = -i|k\rangle \langle l| + i|l\rangle \langle k|$, with $k,l \in \{0,1,2\}$.
Microwave pulses applied resonantly to the 0-1 and 1-2 transitions respectively result in unitaries
$S_{N} = \exp [-i\pi \sigma^{y}_{01}/2(N+1)]$ and $B (\theta_{j}) = \exp (-i\theta_{j} \sigma^{y}_{12}/2)$ ~\cite{supp}. The protocol employs a series of $j=\overline{1,N}$ Ramsey segments, each containing a $B$-pulse with arbitrary strength $\theta_{j}$, overall producing the evolution $U_{N}(\theta_{1}, ... , \theta_{N}) = \prod_{j=1}^{N}[S_{N}B(\theta_{N+1-j})]S_{N} = S_{N}\prod_{j=1}^{N}[B(\theta_{N+1-j})S_{N}]$. Note that the absence of $B$-pulses results in $[S_{N}]^{N+1} = -i \sigma^{y}_{01}+|2\rangle \langle 2|$, acting nontrivially only on the subspace $|0\rangle , |1\rangle$ -- therefore at the end of the sequence  the entire ground-state population is transferred onto the first excited state $|0\rangle \rightarrow |1\rangle$. The goal is to ascertain the presence of $B$-pulses without absorbing them, that is, without creating excitations on level $|2\rangle$ of the transmon.

To understand the interaction-free physics in this setup, consider first a single sequence $N=1$. The transmon is initialized in the ground state $|0\rangle$, which, when acted upon by $S_{1}$ ($\pi/2$ rotation around the $y$-axis in the $\{|0\rangle , |1\rangle \}$ subspace, corresponding to a  $0.5:0.5$ beam-splitter), drives the qubit into a coherent equal-weight superposition state $(|0\rangle + |1\rangle)/\sqrt{2}$. Next, the application of $B(\theta)$ (here we take $\theta_1 \equiv \theta $) and the subsequent application of $S_1$ results in the state $S_{1}B(\theta )S_{1}|0\rangle = \sin^{2}(\theta /4)|0\rangle + \cos^{2}(\theta /4)|1\rangle + (1/\sqrt{2})\sin(\theta /2)|2\rangle$, while if $B(\theta )$ is not present the final state is $|1\rangle$.  By measuring dispersively the state of the transmon and finding it in the state $|0\rangle$, we can successfully ascertain the presence of the $B$ pulse without irreversibly absorbing it. On the other hand, if the transmon is found on $|1\rangle$ we cannot conclude anything, since this is also the result for the situation when the pulse is not present. For the ideal dissipationless case we have $p_{0}(\theta) = \sin^{4}( \theta /4)$, $p_{1}(\theta )= \cos^{4}( \theta /4)$ and $p_{2}(\theta )=(1/2) \sin^2(\theta /2)$. For $\theta =\pi$ this implies that we have $p_{0}(\pi)=25\%$ chance of detecting the $B$-pulse without absorption, leaving $p_{2}(\pi)=50\%$ as the probability of failure due to absorption. 

Our protocol generalizes this concept to a series of $N\geq 1$ sequences, see Fig.~\ref{fig-pulses},  ending with detection by state tomography operators $D_{0}=|0\rangle\langle 0|$, $D_{1}=|1\rangle\langle 1|$, and $D_{2}=|2\rangle\langle 2|$,  which yield the success probability $p_{0} = \langle D_{0} \rangle $, the probability of inconclusive results $p_{1} = \langle D_{1} \rangle$, and the probability of absorption $p_{2}=\langle D_{2} \rangle$. 
In addition, for a given string of $\theta_{j}$'s, as a key figure of merit we define the quantities relevant for the confusion matrix, as employed in standard predictive analytics. The Positive Ratio, PR$=p_{0}/[p_{0}+p_{1}]$, is the fraction of cases where the interaction-free detection of $B$ is achieved \emph{strictly speaking} without irreversible absorption. 
Its counterpart is the Negative Ratio, NR$=p_{1}/(p_{0}+p_{1})$, i.e., the fraction of experiments that are not accompanied by $B$ absorption, but for which we can not ascertain whether a $B$-pulse was present or not. In addition, the so-called interaction-free efficiency is sometimes utilized (see Supplementary Notes 1 and 2 ~\cite{supp}), which for the coherent case reads $\eta_{c} = p_{0}/(p_{0} + p_{2})$. 

We obtain considerable enhancement of the success probabilities and efficiencies when detecting the pulses using this arrangement.

%%%%
\section*{Results}

 As described in the previous section, we use a transmon circuit with a dispersive readout scheme that allows us to measure simultaneously the probabilities $p_0$, $p_1$, and $p_2$. The 0-1 and 1-2 transitions are driven by two pulsed microwave fields, respectively implementing the $S_{N}$ unitaries and the $B$-pulses. Details of simulations and a description of the experimental setup are presented in Methods.
 
 \subsection*{Single $B$-pulse ($N=1$)}
 
 \begin{figure}[ht]
	\centering
	\includegraphics[width=0.8\linewidth]{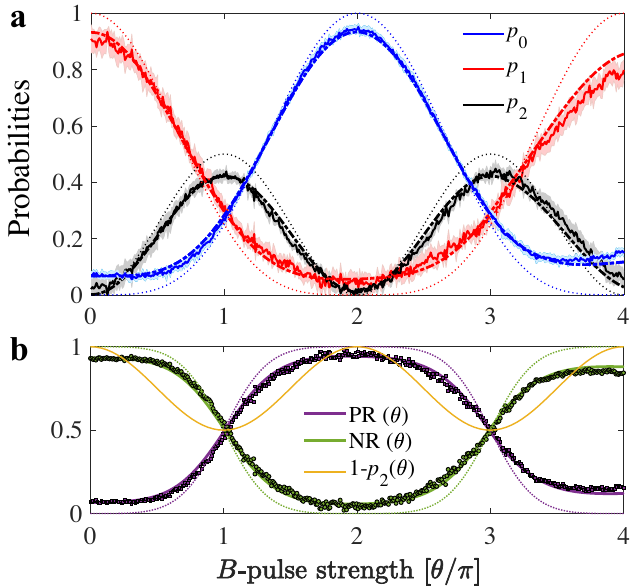}
	\caption{{\bf Probabilities and associated positive/negative ratios for $N=1$.}
	{\bf a} Probabilities vs. strength for a single $B$-pulse in our
three-level system. The experimentally averaged
profiles for the ground state ($p_0$), first excited state ($p_1$) and second 
excited states ($p_2$) are represented by blue, red and black colored continuous lines respectively. The corresponding colored dot-dashed lines are the simulated curves including decoherence and pulse imperfections, while the thin dotted lines show the ideal case. Each experimental curve is accompanied by a shaded region presenting the standard deviation of the mean obtained from 16 replicas of the same experiment. 
{\bf b} Corresponding to each $B$-pulse strength, PR$(\theta )=p_0 (\theta )/[p_0 (\theta )+p_1 (\theta)]$ and NR$(\theta )=p_1 (\theta )/[p_0 (\theta )+p_1 (\theta)]$ obtained from the experiment are shown with purple circular markers and green square markers respectively, closely followed by the simulated purple and green continuous curves. The thin dotted lines represent the respective ideal cases, with no decoherence and without any experimental imperfections. The continuous yellow curve stands for the norm of the system subspace: $p_0(\theta)+p_1(\theta)=1-p_2(\theta)$.}
	\label{fig-1d-1bomb}
\end{figure}
%%%%%%%%%%%%%%%

%%%% 
 \begin{figure}[ht]
	\centering
	\includegraphics[width=0.9\linewidth]{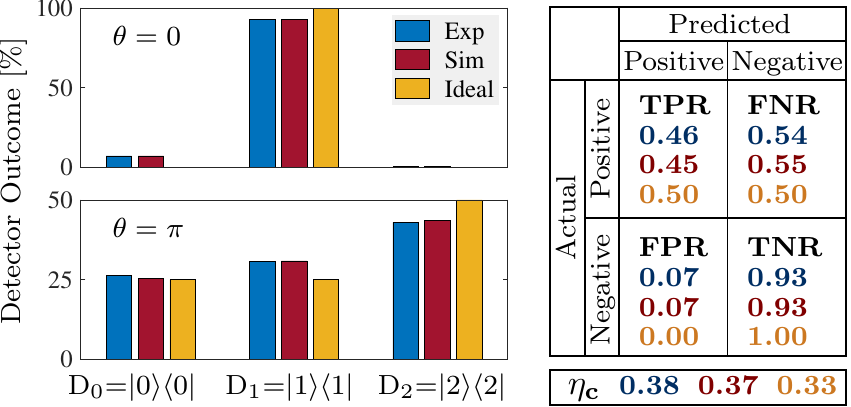}
	\caption{{\bf Histogram of events for $\theta=\pi$ and $N=1$, and the corresponding  confusion matrix and efficiency.} 
		(left panels) Histogram of events recorded by the detectors D$_0$, D$_1$, and D$_2$, which are modeled as projectors. Histograms resulting from the experiments, simulations and ideally expected values are shown in blue, red and yellow colors respectively. The results are obtained from $10^6$ realizations of the experiment, and for $B$-pulse strengths $\theta=0,\pi$. The percentage outcome at D$_0$ corresponds to successful interaction-free detection, D$_2$ represents the number of times the pulse is absorbed, and D$_1$ are the inconclusive instances. (right panel) Confusion matrix and efficiency $\eta_c$ for the detection of $\pi$ pulses showing the experimental 
	(blue), simulated (red) and ideally expected (yellow) values. }
	\label{fig-histogram-1bomb}
\end{figure}
%%%%%%%%%%%%%%%%%%%%%%%%

The $N=1$ case is important since it is the simplest realization of our concept, allowing us to present all the relevant experimental data and the most important figures of merit in a straightforward manner. The main results 
are shown in Fig.~\ref{fig-1d-1bomb} and Fig.~\ref{fig-histogram-1bomb}.

 Fig.~\ref{fig-1d-1bomb}a presents the probabilities $p_{0}$, $p_{1}$, and $p_{2}$ obtained experimentally, as well as a comparison with the simulated values and the ideal case. 
 First, one notices that the results are not invariant under $\theta \rightarrow \theta + 2\pi$, which is intrinsically related to the lack of invariance of spin-1/2  states under $2\pi$ rotations. Indeed, $B (\theta + 2\pi) = \exp (-i\pi \sigma_{12}^{y})B(\theta)$ acts by changing the sign of the probability amplitudes on the subspace $\{|1\rangle , |2\rangle \}$, which subsequently alters the interference pattern after the second beam-splitter unitary. Then, we see that at $\theta=\pi,3\pi$  the experimentally obtained probability for the interaction-free detection is $0.26$; the same would also be expected in the projective case~\cite{Elitzur_1993, supp}.  
 
  From Fig. \ref{fig-1d-1bomb} we also notice that at $\theta=2\pi$ the probability $p_{0}$ reaches a maximum (1 in the ideal case), while $p_{1}$ and $p_{2}$ are minimized (zero in the ideal case).
  This also happens if beam-splitters with $y$-axis rotation angles other than $\pi/2$ are used. It is a situation that has no classical analog: we are able to detect with near certainty a pulse that does not at all change the probabilities. 
  As we will see next, when generalizing this result to $N >1$ pulses, this maximum at $\theta = 2\pi$ extends to form a plateau of large $p_{0}$ values.

We can further characterize the detection capabilities of the $N=1$ protocol by standard predictive analytics methods. In Fig.~\ref{fig-histogram-1bomb} we 
construct the histogram for the presence/absence of a $\theta=\pi$ $B$-pulse and we extract the associated confusion matrix by excluding the cases where the pulse is absorbed. The elements of the confusion matrix are defined by considering an actual positive or negative event (the pulse is either present or not present) and examining what can be predicted about the event based on the detector's response. Using standard terminology in hypothesis testing theory, for our device the elements of the confusion matrix are (see also Supplementary Table 1): when a $\pi$ $B$-pulse has actually been applied, we define the True Positive Ratio $\mathrm{TPR}=p_0(\theta=\pi)/(p_0(\theta=\pi)+p_1(\theta=\pi))=\rm{PR}(\pi)$, which is the fraction of correct detections, and the False Negative Ratio $\mathrm{FNR}= p_1(\theta=\pi)/(p_0(\theta=\pi)+p_1(\theta=\pi))=\rm{NR}(\pi)$, which is the fraction of inconclusive events. When the pulse is not applied, we have the False Positive Ratio $\mathrm{FPR} = p_0(\theta=0)/(p_0(\theta=0)+p_1(\theta=0))=\rm{PR}(0)$, which is the fraction of times we would wrongly predict that the pulse was applied, and its complementary True Negative Ratio $\mathrm{TNR}= p_1(\theta=0)/(p_0(\theta=0)+p_1(\theta=0))=\rm{NR}(0)$, which are the cases where we cannot predict anything.  Finally, for the efficiency we obtain $\eta_{c} (\theta =\pi) = 0.33$ (refer to Supplementary Fig.~2
for other values).
The experimental results in Fig.~\ref{fig-histogram-1bomb} are well reproduced by simulations and close enough to the ideal values.

\subsection*{Two consecutive $B$-pulses ($N=2$)}

\begin{figure}[ht]
	\centering
	\includegraphics[width=0.9\linewidth]{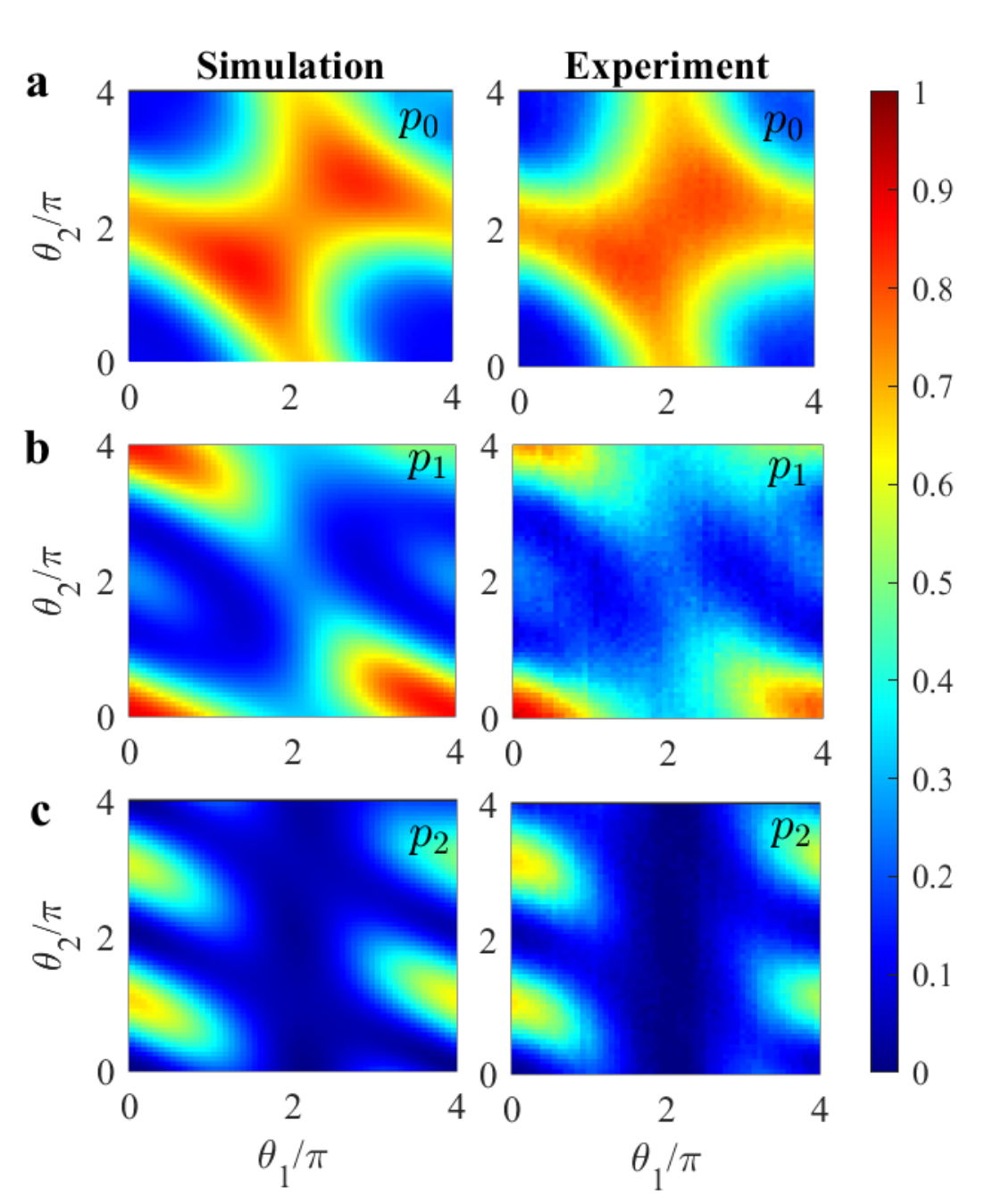}
	\caption{{\bf Probabilities for the $N=2$ case.}
		2D probability maps for the {\bf a} ground state ($p_0$), {\bf b} first excited state ($p_1$), and {\bf c} second excited state ($p_2$) as a function of $B$-pulse strengths $\theta_1$ and $\theta_2$.} 
	\label{fig:012}
\end{figure}

\begin{figure}[h]
	\centering
	\includegraphics[width=0.9\linewidth]{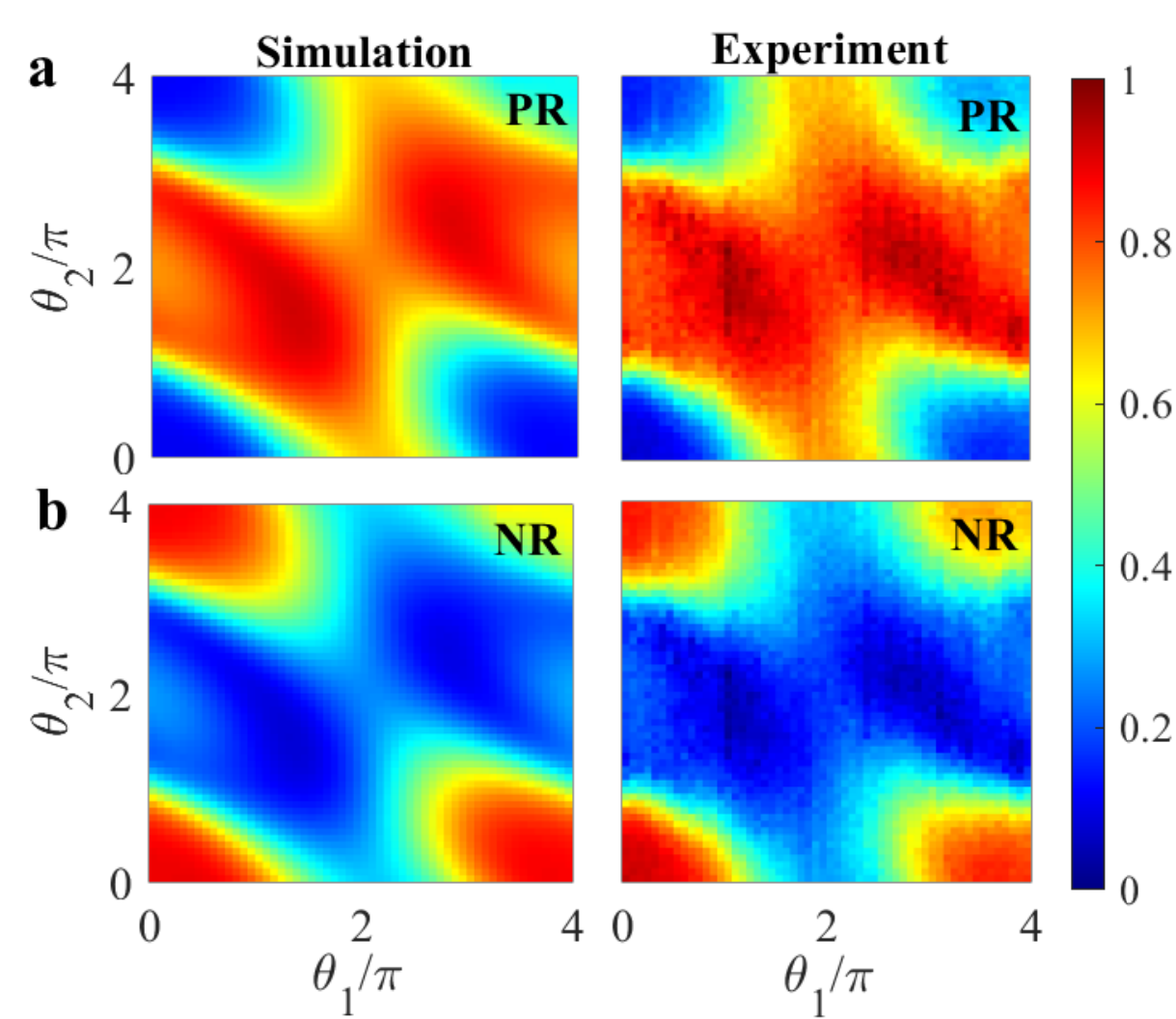}
	\caption{{\bf Positive and negative ratios for the $N=2$ case.}
		Simulated and experimental 2D maps for the {\bf a} Positive Ratio $\mathrm{PR}(\theta_1, \theta_2)$ and for the {\bf b} Negative Ratio $\mathrm{NR}(\theta_1, \theta_2)$ as a function of $\theta_1$ and $\theta_2$.}
	\label{fig:012PRNR}
\end{figure}

Next, we use our superconducting circuit to realize the coherent interaction-free detection of $N=2$ pulses. 
The sequence of operations consists of two independent $B$-pulses of strengths $\theta_1$ and $\theta_2$ sandwiched between three beam-splitter unitaries. 
In this case the coherent protocol already becomes fundamentally different from the projective one. Further, for $N=2$, one can conveniently study all possible combinations of the pair of $B$-pulses whose strengths $\theta_1, \theta_2 \in [0, 4\pi]$ can be varied independently. This also allows us to study new situations, such as the absence of one of the $B$-pulses. 

The experimental and the simulated 
results for the probabilities associated with the ground state, the first excited state and the second excited 
state as functions of $\theta_1$ and $\theta_2$ are shown in Fig.~\ref{fig:012}a--c, respectively. 
The Positive Ratio PR$(\theta_1, \theta_2)=p_0(\theta_1, \theta_2)/(p_0(\theta_1, \theta_2)+p_1(\theta_1, \theta_2))$ and the Negative Ratio NR$(\theta_1, \theta_2)=p_1(\theta_1, \theta_2)/(p_0(\theta_1, \theta_2)+p_1(\theta_1, \theta_2))$ as functions of $\theta_1$ and $\theta_2$ are shown in Fig.\ref{fig:012PRNR}. Similar to the $N=1$ case, the PR and NR can be used to construct the confusion matrix for any combination of $\theta_{1}$ and $\theta_{2}$ values. For the efficiency we obtain $\eta_{c} (\theta_{1} =\pi , \theta_{2} =\pi ) =  0.81$ (refer to Supplementary Fig.~3 for other values). 
The experimental and simulated results are in very good agreement with each other, demonstrating control of the system over the full range of the two $\theta$-parameters.

To understand the difference between the coherent and the projective protocol, let us look at the case $\theta_1=\theta_2 = \pi$. The projective protocol, if the first pulse is not absorbed, 
produces the state $|0\rangle$ at the input of the second beam-splitter unitary (see Supplementary Note 2)~\cite{supp}. As a result, the second Ramsey sequence provides another round of monitoring the pulse, though this is essentially only a repetition of the first. In contrast, in the coherent protocol the input to the second beam-splitter unitary is a superposition of $|0\rangle$ and $|2\rangle$. The second monitoring of the pulse retains the amplitude of $|2\rangle$ in a coherent way, resulting in a higher probability of success.
This unexpected effect can be seen by a straightforward calculation for the ideal case and $\theta_1 = \theta_2 = \pi$, which yields probabilities $p_{0}=0.8091$, $p_{1}= 0.0034$, $p_{2}=0.1875$, and PR$=0.99$; whereas, the equivalent respective figures for the projective case are 0.4219, 0.1406, 0.4375, and 0.75.

\subsection*{Multiple consecutive $B$-pulses ($N>2$)}

Next, we use our superconducting circuit to realize the coherent interaction-free detection of $N>2$ pulses, where we observe even 
more efficient coherent accumulation of the amplitude probabilities on the state $|0\rangle$ under successive interactions with the $B$-pulse and applications of Ramsey ~$S_{N}$\cite{supp}.  

\begin{figure*}[ht]
	\centering
\includegraphics[width=0.95\linewidth]{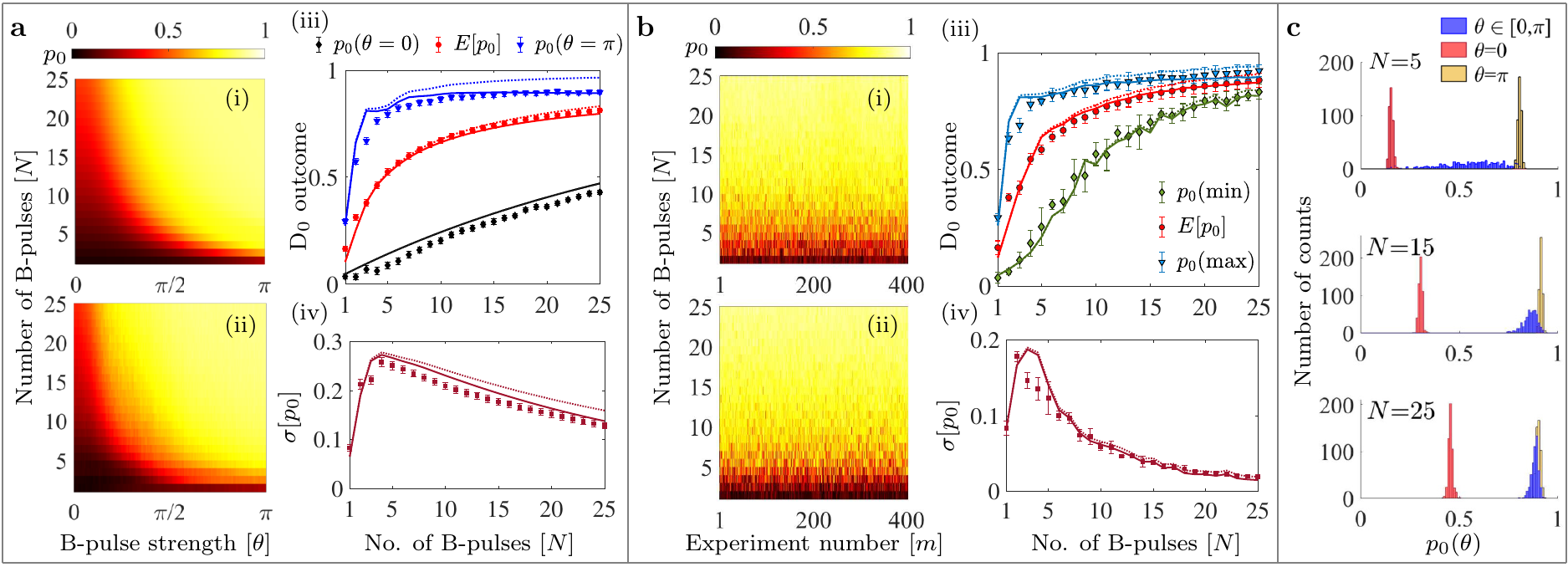}
	\caption{{\bf Results for the  D$_0$ outcome in the case of multiple Ramsey sequences with $B$-pulses 
	$N\in [1,25]$ of varying strengths, $\theta_{j,m} \in [0,\pi]$
	with $j=\overline{1,N}$ and $m$ indexing the experimental realization
	for a given $N$.}
	{\bf a} Plots for identical $B$-pulses  $\theta_{j,m}=\theta=m\pi/\mathcal{M}$ for a given $N$: 
	(i) Simulated and (ii) 
	experimentally obtained maps.  
	(iii) Values of $p_0$ at $B$-pulse strengths $\theta=0,\pi$ and mean $p_0$ (E$[p_0]$)
	versus $N$ with 
	markers with error bars showing the experimental results and the 
	corresponding continuous lines obtained from the simulations. Red circular markers present the mean value $E[p_0]$, black diamond markers correspond to the case with no $B$-pulses ($\theta=0$) and data points with blue triangular markers stand for the case of maximum $B$-pulse strengths ($\theta=\pi$). (iv) Standard deviation evaluated for a given $N$ versus $N$. 
	{\bf b} Plots for arbitrarily chosen
	$\theta_{j,m} \in [0,\pi]$: 
	(i) Simulated and (ii) experimental data for $p_0$
	as a function of $N$ and $m$. (iii) Simulated (continuous lines) and experimental (markers with error bars) of mean ($E[p_0]$ in red) and extremum values ($p_0$(min) in green and $p_0$(max) in blue). (iv) Standard deviation of $p_0$ versus $N$. 
	Dotted curves in all the plots are simulations without the inclusion of depolarization.
	Error bars in all the plots correspond to standard deviation of the measured quantities with respect to their respective mean values, obtained from four repetitions of the full experiment.
	{\bf c} Histogram of the experimental $D_{0}$ counts for various system sizes ($N = 5,15,25$) with $B$-pulses of arbitrary strengths, 
	$\theta \in [0, \pi]$ (in blue) compared with those of $B$-pulse strengths: $\theta = \pi$ (in yellow) and $\theta = 0$ (no $B$-pulse) in red.
	Clearly as the system size increases, the strengths of the $B$-pulses become less significant and approach the clustering near the maximal $p_0$, which is a signature of the highly efficient interaction-free detection.}
	\label{fig:multipleBombs}
\end{figure*}

In these experiments we use both equal-strength pulses $\theta_j = \theta$ and pulses with randomly-chosen $\theta_j\in \{0, \pi\}$, $j=\overline{1,N}$, while the  beam-splitter unitary is a $\pi/(N+1)$ rotation around the $y$ axis in the $\{|0\rangle,|1\rangle\}$ subspace. To recall, in the absence of the $B$-pulses we have $[S_{N}]^{N+1}$ and in the presence of the $B$-pulses we have $S_{N}\prod_{j=1}^{N}[B(\theta_{N+1-j})S_{N}]$. The results are presented in Fig. \ref{fig:multipleBombs}.
Due to the multidimensional nature of these experiments we focus here on $p_0$; other possible figures of merit are presented in Supplementary Information Note 2(c).

%\onecolumngrid

%\twocolumngrid

The large-$N$  experimental sequences  have a significant time cost with the worst case of $25$ $B$-pulses corresponding to $4.3\;\mu$s, which is even longer than the relaxation time $\Gamma_{10}^{-1} = 3.4\;\mu$s (see Methods for details). 
 Thus, in addition to the standard three-level Lindblad 
master equation~\cite{open-quantum-system-2002, Kumar_2016}, in  order to accurately model the system we may include a depolarizing channel $\rho(t) \rightarrow (1-\epsilon) \rho(t) + \epsilon \mathbb{I}_{3}/3$~\cite{nielsen-book-2002} (see Methods). Here we assume that the imperfections in the $1-2$ drive results in mixing of the qutrit state; hence the parameter $\epsilon$ is taken as directly proportional to the pulse amplitude, given by $\epsilon[\theta] = 1.8\times 10^{-3} \times \theta/\pi$.
 This choice of model fits our experimental data very well as shown in Fig.~\ref{fig:multipleBombs}, where continuous lines correspond to the simulation including the depolarizing channel and dotted lines correspond to the simulation without the depolarizing channel. As expected, the overall effect of depolarization is more prominent for a larger number of $B$-pulses and for large $\theta$. In all of these plots, experimental results are shown by markers with experimental error bars (standard deviation about the mean by four repetitions of the same experiment). Small deviations of the experimental values from the ideal results are due to decoherence and pulse errors. Larger values of $p_0$ correspond to a higher probability of interaction-free detection. We have verified numerically that with increasing $N$, $p_0$ increases, approaching $1$ in an ideal case.

 In the case of equal-strength pulses, for each $N$,
 we perform a total of $\mathcal{M}$ experiments, with the $B$-pulse strength
 varying linearly with the experiment number as: 
 $\theta=\theta_{j,m}=m\pi/\mathcal{M}$
 with labels: $j=\overline{1,N}$ and $m=\overline{1,\mathcal{M}}$ such that $\theta \in [0,\pi]$. The results for the overall success probability $p_0$ are shown in Fig.~\ref{fig:multipleBombs}a, for various numbers $N\in [1,25]$ of $B$-pulses and $\mathcal{M}=180$. Simulated and experimental $p_0$ values are shown as surface plots in parts (i) 
and (ii) respectively. 
 
Interestingly, with increasing number of $B$-pulses, the final 
$p_0$ is independent of the $B$-pulse strength ($\theta$), and has a tendency to reach
large values. As anticipated, a plateau characterized by high values is formed, which is the extension to smaller $\theta$'s of the maximum seen in the $N=1$ case around $\theta =2\pi$. This is also clearly reflected from the plot in Fig.~\ref{fig:multipleBombs}a(iii) showing the mean value of $p_0$ ($E[p_0]$ in red) resulting from experiments with different $B$-pulse strengths versus the number of Ramsey sequences. The `no $B$-pulse' situation is shown with black square markers and that of maximum $B$-pulse strength is shown with blue triangular markers, where the increase in $p_0 (\theta=0)$ with $N$ and lower values of $p_0 (\theta=\pi)$ is due to the decoherence. It is clear from the three curves that $E[p_0]$ tends to approach the higher limiting values, which is attributed to the larger plateau of high $p_0$ values with increasing $N$ (see Supplementary Fig.~6)\cite{supp}.  
As a direct consequence of the plateau formation, the minimum value of $\theta$ that gives rise to near maximal $p_0$ is much smaller than $\pi$ for large $N$.
The standard deviation of the $p_0$ distribution versus $N$ is shown in Fig.~\ref{fig:multipleBombs}a(iv). Each of these experimental values are accompanied by simulations, demonstrating quite close agreement. A comparison (see Supplementary Notes 2 and 3)\cite{supp} with the projective case - for which exact analytical results are available - demonstrates the advantage of the coherent protocol for all values of $N$.

We also study the case of randomly-chosen $\theta_j\in \{0, \pi\}$, $j=\overline{1,N}$, with results 
shown in Fig.~\ref{fig:multipleBombs}b. Panels (i), (ii) present surface maps of the simulated and experimental $p_0$ versus $N$ and $m$, where $\mathcal{M}=400$.
Experimental and simulated mean- $E[p_0]$, minimum- $p_0^{\rm (min)}$, and maximum- $p_0^{\rm (max)}$ values obtained from this distribution are shown in panel (iii) with markers and continuous curves respectively. The standard deviation $\sigma[p_{0}]=\sqrt{\mathrm{E}[(p_{0}-\mathrm{E}[p_0])^2]}$ of $p_0$  versus $N$ is shown in part (iv). 
Again, we observe that the mean value of $p_{0}$ increases with $N$, while the standard deviation of repeated measurements decreases with N. Thus, for a large N, the $B$-pulse strength does not matter anymore, and we obtain a  highly effective interaction-free detection. 
Surprisingly, the case with random $B$-pulse strengths appears to outperform the case with identical $B$-pulses. Comparing parts 
a(iii) and b(iii) of Fig.~\ref{fig:multipleBombs}, the success probability of 
the coherent interaction-free detection in the worst case (green curve) for
random $B$-pulse strengths is already high enough, with a maximum value (for $N=25$) of $0.83 \pm 0.03$ (experiment) and $0.82$ (simulation), close to the mean values $E[p_0]=0.88 \pm 0.03$ (experimental) and $E[p_0]=0.87$ (simulated).
On the other hand, in the case of identical $B$-pulses, the mean values for $N=25$ are only
$E[p_0]=0.81 \pm 0.01$ (experiment) and $E[p_0]=0.80$ (simulation), even slightly below the worst-case scenario with random pulses.
Also, especially at large $N$'s, the standard deviation about the mean value of the distribution 
is much lower in the case of random $B$-pulses as opposed to the identical $B$-pulses case, which is clear upon comparison of Fig.~\ref{fig:multipleBombs}a(iv) and b(iv).
Thus, an adversarial attempt to randomize the $B$-pulse strengths in order to evade detection has, surprisingly, the opposite effect, improving the interaction-free coherent detection.

In Fig.~\ref{fig:multipleBombs}c we provide a histogram representation of the $p_0$ distributions for $N=5,15,25$. The distribution  in red in all three cases
corresponds to $\theta_j=\theta=0$ -- and hence lie at the lower limit of $p_0$ range, 
while the distribution in yellow represents the case  $\theta_j=\theta=\pi$ and lies close to
the upper limit. The interesting part is the distribution in blue with arbitrarily chosen $B$-pulse strengths
$\theta_j=\theta \in [0,\pi]$, which moves towards the right side and tends to squeeze with increasing $N$.
The same idea is conveyed by the increasing mean value ($E[p_0]$) and decreasing standard deviation with $N$ as discussed earlier.

Finally, as another figure of merit for the protocol, we can obtain PR$(\theta)$ and NR$(\theta)$ for $B$-pulses with equal strengths $\theta_{j,m} = \theta \in [0,\pi]$ for each $N \in [1,25]$. The detailed surface maps presenting the ideal case (without decoherence), and the simulated and experimentally obtained values for PR$(\theta)$ and NR$(\theta)$ at various $N$ are shown in Supplementary Fig.~5. Similar to the previous cases, these can be used to define the elements of the confusion matrix, for example TPR = PR($\pi$), FPR = PR($0$), etc.
 	We find that at large N the positive ratio  reaches high values for a wide range of $\theta$'s, altoghether forming a plateau of stable and high-confidence interaction-free detection. 
 	 Correspondingly, a wide region of low NR$(\theta)$ values are obtained.
For example, from the experimental data, for N = $5, 15, 25$ the value PR$(\theta)=0.90$ is reached at $\theta=0.54\pi, 0.32\pi, 0.18\pi$ respectively, going up to $\approx 0.95$ at $\theta=\pi$. The corresponding values of the efficiency $\eta_c$ for the same $N$ and $\theta$ combinations are $0.67$, $0.81$ and $0.81$ respectively, see also Supplementary Fig.~6.

%%%%%%%%%%%%%%%%%%%%%%%%%%%%%%%%%%%%%%%%%%%%%%%%%%%

\section*{Discussion}

In our protocol quantum coherence serves as a resource, yielding a significantly high detection success probability. The enhancement can be understood as the coherent accumulation of amplitude probabilities on the state $|0\rangle$ under successive interactions with the $B$-pulse and applications of Ramsey $S_{N}$ (see Supplementary Note 3)~\cite{supp}, by making use of the full 3-dimensional Hilbert space at each step.
In contrast, the projective protocol \cite{Kwiat_1995,Kwiat_1999} employs the quantum Zeno effect to confine the dynamics in the ${|0\rangle , |1\rangle}$ subspace after each interaction with the pulse. Thus, it extracts which-way information about the presence or absence of the pulse at each step of the protocol.

To gain more insight into the functioning of our protocol, consider the case of uniform $B$ $\pi$-pulses.
We have verified numerically that at large values of $N$ the following approximate relation holds
	\begin{equation} 	
		U_{N}(\theta_{1}=\pi, ..., \theta_{N}=\pi) = [S_{N}B(\pi)]^{N}S_{N}  \overset{\scriptscriptstyle{N\gg1}}{\approx} |0\rangle \langle 0| + \left(-i \sigma_{12}^{y}\right)^{N}  \nonumber
	\end{equation}
We can also provide a consistency argument for this relation: since we are dealing with $\pi$ pulses only, we have $B(\pi) = |0\rangle \langle 0| - i \sigma_{12}^{y}$, and since $N\gg 1$ we can write also $S_{N+1}\approx \mathbb{I}_{3}$. Then, assuming the above expression, we can estimate $U_{N+1}(\theta_{1}=\pi, ..., \theta_{N+1}=\pi) \approx U_{N}(\theta_{1}=\pi, ..., \theta_{N}=\pi)B(\pi)\mathbb{I}_{3} = |0\rangle \langle 0| + \left(-i \sigma_{12}^{y}\right)^{N+1}$. 
Thus, if we start from the ground state, the dynamics tend to stabilize this state at large $N$, which results in the appearance of plateaus of near-unity $p_{0}$ in Fig.~\ref{fig:multipleBombs} a. This is in some sense the closest counterpart of the approximation $\left[\cos(\pi /2(N+1))\right]^{2(N+1)} \overset{\scriptscriptstyle{N\gg1}}{\approx} 1$, which is crucial for establishing a large detection in the standard projective case (see also Supplementary Note 2).

In the experimental realization of projective interaction-free measurements, as done with bulk optics \cite{Kwiat_1999} or waveguide circuits \cite{Ma2014}, the maximum experimental efficiencies obtained are 0.73 and 0.63 resepectively, both obtained for $N=9$. For larger $N$'s it is observed that the efficiency decreases due to losses. By contrast, in our case the efficiency for $N=9$ is $\eta_{c} (\theta =\pi)=0.89$ and it increases further as $N$ gets larger, reaching 0.96 at $N=20$ (see also Supplementary Fig.~6).
Our protocol also compares favorably with other realizations of microwave photon detection, based for example on Raman processes \cite{Inomata_2016}, or on cavity-assisted conditional gates \cite{Nakamura2018,Wallraff2018}. The dark count rate, which is the number of counts per unit time in the absence of a pulse, can be obtained from $\mathrm{FPR} \approx p_{0}(\theta =0)$ divided by the sensing time: we obtain 0.1 counts/$\mu$s. This can be further improved without affecting the true positives by reducing the decoherence and the effective qubit temperature at the beginning of the protocol, for example by using active reset. The experimentally-demonstrated detection bandwidth of our system is given by the inverse minimum duration of the $B$-pulses used in the experiment; e.g., for the 56 ns pulses this corresponds to a 18 MHz bandwidth.

\begin{figure}[h]
 	\centering
 	\includegraphics[width=1\linewidth]{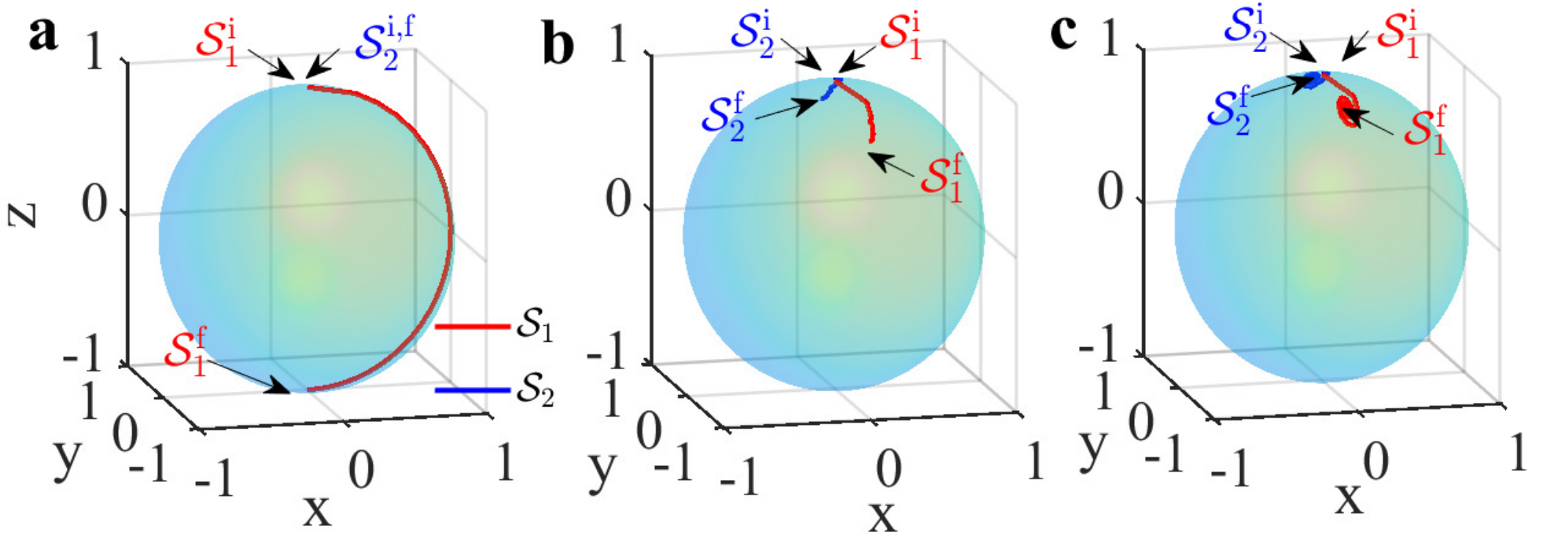}
 	\caption{{\bf Majorana representation.} Averaged Majorana trajectories followed by the three-level system for $N=25$ in the case of {\bf a} no $B$-pulse,  $B$-pulses with {\bf b} equal strengths, and {\bf c} with randomly chosen strengths in the range $\left[ 0, \pi \right]$.
 	Trajectories of the Majorana stars $\mathcal{S}_1$ and $\mathcal{S}_2$ are shown in red and blue colors respectively, where $\mathcal{S}_{1,2}^{\rm i}$ marks the initial state and $\mathcal{S}_{1,2}^{\rm f}$ correspond to the final state of the three-level system on the Majorana sphere.
 	}
 	\label{fig-majorana25}
 \end{figure}

The coherent interaction-free protocol can also be represented geometrically on the unit 2-sphere. In the Majorana representation~\cite{majorana-nc-1932}, a three-level system is represented by two points $\mathcal{S}_{1}(x,y,z)$ and $\mathcal{S}_{2}(x,y,z)$ -- called Majorana stars -- on the surface of this sphere ~\cite{Dogra2020}. 
	 In our protocol, the system is initialized in the state $|0\rangle$, which corresponds to both Majorana stars residing at the North Pole, $\mathcal{S}_{1,2}^{\rm i}(0,0,0)$. In the absence of $B$-pulses, the protocol ends with one star at the North Pole and the other at the South Pole. In the presence of $B$-pulses with $\theta_j=\pi$, we find that both stars are located in the northern hemisphere for N$\geq2$, and they tend to get closer and closer to the North Pole with increasing $N$ (see also Supplementary Note 6)~\cite{supp}.
 	 To illustrate this, in Fig.~\ref{fig-majorana25}a--c we present the resulting trajectories of the Majorana stars ($\mathcal{S}_1$ in red and $\mathcal{S}_2$ in blue) for the case of no $B$-pulse, $B$-pulses with equal strengths, and $B$-pulses with randomly chosen strengths respectively. Here we took $N=25$, such that each Majorana trajectory consists of $26$ points; the initial and final stars of the trajectories are labelled as $\mathcal{S}_{1,2}^{\rm i}$ and $\mathcal{S}_{1,2}^{\rm f}$ respectively. The trajectories correspond to the average states obtained from 400 repetitions of the protocol with varying $B$-pulse strengths (as discussed in the previous section). The presence of both Majorana stars in the vicinity of the North Pole on the sphere serves as a sensitive  geometrical signature of the interaction-free detection of the $B$-pulses. There is a clear difference between the situation of no $B$-pulse, where one Majorana star is at the North Pole (0,0,1) and the other at the South Pole (0,0,-1), as compared to the presence of the $B$-pulse, shown in Fig.~\ref{fig-majorana25}b and c, where both $\mathcal{S}_1$ and $\mathcal{S}_2$ end up close to the North Pole. Comparing Fig.~\ref{fig-majorana25}b and c, we find that the z-coordinates of the final Majorana stars in the case of equal $B$-pulse strengths is $0.7381$, while the minimum value of the z-coordinate reached in the case of randomly chosen $B$-pulse strengths is $0.7863$. Clearly, in the case of randomly chosen $B$-pulse strengths the respective Majorana trajectories are confined closer to the North Pole, confirming the results from the previous section.
  
We point out that these results can be extended in various directions. For example, they can be applied for the non-invasive monitoring of microwave currents and pulses, which is an open problem in quantum simulation \cite{Geier_2021}. They provide a proof of concept for a photon detector, conceptually and practically different from  realizations based on other principles, that can be further optimized.
Our protocol works also when the $B$-pulse is a Fock state and it can be utilized to assess non-destructively the presence of photons stored in superconducting cavities (see Supplementary Note 1)~\cite{supp}. This can be utilized for axion detection, where the generation of a photon is expected to be a rare event. Here also the existing detectors have a high dark count rate; thus, one can increase the confidence level by assessing its presence first non-destructively and then confirming it by more conventional means.

% \section*{Conclusions} 
% This heading is commented as suggested by author guidelines

In conclusion, we proposed a coherent interaction-free process for the detection of microwave pulses and we realized it experimentally with a superconducting quantum circuit. For the case of a single pulse with strength $\theta =\pi$, we 
obtain an interaction-free detection probability of $0.26$.
Further, we emulated multiple Ramsey sequences and we obtained a highly efficient interaction-free 
detection of the $B$-pulse. We observed that for a large number of sequences a detection probability approaching unity is obtained irrespective to the strength of the pulses, and, surprisingly, this probability is even higher when the pulses have  random strength.

%%%%%%%%%%%%%%%%%%%%%%%%%%%%%%%%%%%%%%%%%%%%%%%%%%%%%  
  \section*{Methods}
  \subsection*{Experimental Setup}

\begin{figure}[h]
	\centering
	\includegraphics[width=0.8\linewidth]{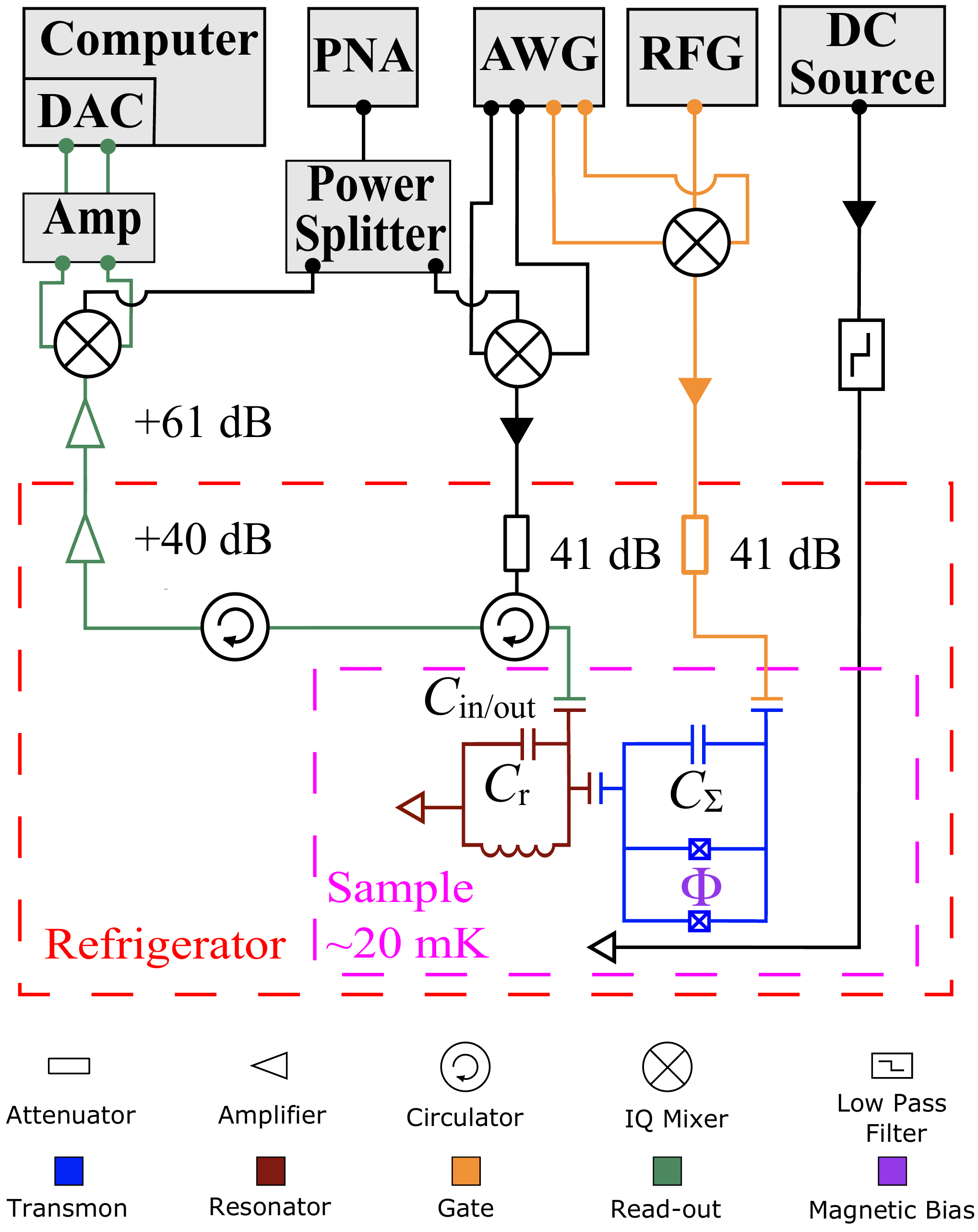}
	\caption{{\bf Experimental setup.}
	Schematic of the experimental setup used in this work, including the transmon circuit’s integration with the dilution refrigerator and microwave electronics.}
	\label{fig:setup}
\end{figure}

A schematic of the setup is shown in Fig.~\ref{fig:setup}. The sample is mounted in a dilution refrigerator via a sample holder which is thermally anchored to the mixing chamber. There are several lines that connect our sample to the external circuitry: the microwave gate line which delivers the microwave drive pulses to the transmon, a flux-bias line which provides a constant DC magnetic field, and the measurement line which is capacitively coupled to the readout resonator via an input/output capacitor. The flux-bias line sends a current near the SQUID loop, which induces a magnetic flux and thus enables the transmon transition frequency to be tuned. To reduce the sensitivity of the device to charge noise, the SQUID loop is shunted by a large capacitance \cite{You_2007, Koch_2007, Barends_2013} denoted by $C_{\rm \Sigma}$ in Fig.~\ref{fig:setup}. The transmission line is used to probe the resonator by sending microwave pulses or continuous signals into it.

The drive pulses used to realize the beam-splitter unitaries and the $B$-pulses 
have super-Gaussian envelopes ($ \propto \exp{[-(t/\tau)^4/2]}$) with the following time-dependence: 
\begin{equation}
\Omega(t) = \Omega_{0} \exp{\left[-\frac{1}{2}\left(\frac{t}{\tau}\right)^4\right]}\;, \label{eq:omegat}
\end{equation}
where $\Omega^{(S_{N})}_{0} =\pi/[(N+1)\int^{\tau_{c}}_{-\tau_{c}} \exp{[-(t/\tau)^4/2}] \mathrm{d} t]$ for beam-splitters and $\Omega_{0}(\theta) = \theta/\int^{\tau_{c}}_{-\tau_{c}} \exp{[-(t/\tau)^4/2]} \mathrm{d} t$ for the $B$-pulses. Thus, the effective pulse area is determined by $\int^{\tau_{c}}_{-\tau_{c}} \exp{[-(t/\tau)^4/2]} \mathrm{d} t$, where  
$\pm \tau_{c}$ are the start and the end points of the drive pulse (the points where the pulse is truncated) and $\tau$ is a time constant. In our experiments $\tau=14$ ns and $\tau_{c} = 2 \tau =  28$ ns, which corresponds to a total pulse length of 56 ns and an effective pulse area $\int^{\tau_{c}}_{-\tau_{c}} \exp{[-(t/\tau)^4/2]} \mathrm{d} t =30.18$ ns. The amplitude $\Omega_{0}$ is determined from Rabi oscillations measurements varying the amplitude of the transmon drive pulse and its frequency while keeping the pulse duration fixed. The variation of the pulse amplitude is achieved using I and Q waveform amplitudes from our arbitrary waveform generator (AWG), which are mixed in an IQ mixer with the LO tone generated by a continous microwave generator (AWG). We utilize a homodyne detection scheme for determining the state of the transmon. A microwave source (PNA) provides a continuous signal at the LO frequency for our readout pulse as well as that for the demodulated reflected signal from the resonator. As such, a power splitter is employed to halve this signal, where one part is sent to the LO port of an IQ mixer which modulates a probe pulse with readout rectangular envelopes from the I and Q quadratures generated by the AWG.  The other part is sent to an IQ mixer which demodulates the signal reflected back from the resonator. After demodulation the quadratures of this mixer are amplified and subsequently digitized and recorded via our data acquisition card (DAC).

\subsection*{Decoherence model and numerical simulations}

In the rotating wave approximation (RWA), the transmon Hamiltonian in the three-level truncation is
\begin{equation}
    \begin{split}
    H(t) = \frac{\hbar}{2}\Big{[}\Omega_{01}(t)e^{i\phi_{01}}|0\rangle \langle 1| + \Omega_{01}(t)e^{-i\phi_{01}} |1\rangle \langle 0|+ 2\delta_{01}|1\rangle \langle 1|\Big{]} \\ + \frac{\hbar}{2}\Big{[}\Omega_{12}(t)e^{i\phi_{12}}|1\rangle \langle 2| +  \Omega_{12}(t)e^{-i\phi_{12}}|2\rangle \langle 1| + 2(\delta_{01}+\delta_{12})|2\rangle \langle 2|\Big{]}\;,
    \end{split}
\end{equation}
where the drive amplitudes follow the form as per Eq.~(\ref{eq:omegat}), and are denoted as $\Omega_{01}(t)$ and $\Omega_{12}(t)$ for the $|0\rangle-|1\rangle$ and $|1\rangle-|2\rangle$ transitions respectively, carrying the respective phase factors $e^{\pm i\phi_{01}}$ and $e^{\pm i\phi_{12}}$~\cite{Kumar_2016}.
With the notation $\sigma_{kl}=|k\rangle \langle l|$, and assuming resonance $\delta_{01}=\delta_{12}=0$, the Hamiltonian reads
\begin{equation}
		H(t) = \frac{\hbar\Omega_{01}(t)}{2}e^{i\phi_{01}}\sigma_{01} +  \frac{\hbar \Omega_{12}(t)}{2}e^{i\phi_{12}}\sigma_{12} + \mathrm{h.c.} \label{supp-eq3}
\end{equation}

To introduce dissipation, we use the standard Lindblad master equation, 
where $D[L]\rho = L\rho L^{\dagger} - \frac{1}{2}\{L^{\dagger}L ,\rho \}$ is the Lindblad super operator and $L$ is the jump operator applied to the density matrix $\rho$. For our three-level system we have (see e.g. \cite{Li2012,Guzik2011})
\begin{equation*}
    \dot{\rho} = -\frac{i}{\hbar}[H,\rho] + \sum_{k,l = 0,1,2}^{k\neq l}  \Gamma_{k\rightarrow l}D[\sigma_{lk}]\rho +  \sum_{k = 0,1,2}\frac{\Gamma^{\phi}_{k}}{2}D[\sigma_{kk}]\rho, %\label{eq:Lind1}
\end{equation*}
where $\Gamma_{k\rightarrow l}$  is the excitation/decay rate between states $|k\rangle$ and $|l\rangle$, and 
$\Gamma_{k}^{\phi}$ is the dephasing rate associated with level $k$. The operators $\sigma_{lk} =|l\rangle \langle k|$ with $k >l$ are lowering operators and those with $k<l$ are raising operators corresponding to the transition $lk$. The Lindblad dephasing operators act only on the off-diagonal matrix elements, while the relaxation operators act on both the diagonal and off-diagonal matrix elements.
However, since we operate on transitions, the individual dephasing rates $\Gamma_{k}^{\phi}$ cannot be determined directly from experiments. Instead, we can rewrite the equation above in a form that involves only pairs of levels~\cite{Kumar_2016}

\begin{eqnarray*}
	\dot{\rho} &=& -\frac{i}{\hbar}[H,\rho] + \Gamma_{2\rightarrow 1}\rho_{22}(\sigma_{11}-\sigma_{22}) + \Gamma_{1\rightarrow 0}\rho_{11}(\sigma_{00}-\sigma_{11}) \\
	& &+ \Gamma_{1\rightarrow 2}\rho_{11}(\sigma_{22}-\sigma_{11}) + \Gamma_{0\rightarrow 1}\rho_{00}(\sigma_{11}-\sigma_{00}) \\ 
	& & -\sum_{k,l = 0,1,2}^{k\neq l} \gamma_{kl}\rho_{kl}\sigma_{kl}, \nonumber
\end{eqnarray*}
where the relaxation rates satisfy the detailed balance condition $\Gamma_{k\rightarrow l}= e^{-\hbar \omega_{kl}/\rm{k_{\rm B}T}}\Gamma_{l\rightarrow k}$ (with $l>k$) at a temperature T with $k_{\rm B}$ being the Boltzmann constant and $\hbar \omega_{kl}$ being the energy level spacing between the $k^{th}$ and $l^{th}$ 
levels. By introducing the occupation numbers $n_{kl}=1/[\exp(-\hbar \omega_{kl}/k_{\rm B}T)-1]$, the rates  $\Gamma_{k\rightarrow l}$ can be expressed in terms of the zero-temperature decay rates $\Gamma_{lk}$ (with $l>k$) as $\Gamma_{k\rightarrow l} = n_{kl}\Gamma_{lk}$  ($l>k$)  and 
$\Gamma_{l\rightarrow k} = (n_{kl}+1)\Gamma_{lk}$  ($l>k$).
It is clear from this decoherence model that the relaxation rates $\Gamma_{k\rightarrow l}$ for $k<l$ are significant only at higher temperatures of several tens of mK, which lead to transitions from lower to higher energy levels. 
The decay rates for the off-diagonal matrix elements are $\gamma_{10} = \gamma_{01} = (\Gamma_{1\rightarrow 0}+\Gamma_{0\rightarrow 1})/2 + \Gamma_{10}^{\phi}$, $\gamma_{21} = \gamma_{12} = (\Gamma_{1\rightarrow 2}+\Gamma_{2\rightarrow 1})/2 + \Gamma_{21}^{\phi}$, and $\gamma_{20} = \gamma_{02} = (\Gamma_{1\rightarrow 0} + \Gamma_{2\rightarrow 1} + \Gamma_{0\rightarrow 1} + \Gamma_{1\rightarrow 2})/2 + \Gamma_{20}^{\phi}$. Here we define the dephasing rates associated with each transition as  $\Gamma^{\phi}_{kl}=\Gamma^{\phi}_{lk}=
(\Gamma^{\phi}_{k}+\Gamma^{\phi}_{l})/2$. Note that the off-diagonal decay of the matrix elements $\rho_{kl}$ due to dephasing can be understood as resulting from $\mathbb{I}_{kl}D[\sigma^{z}_{kl}]\rho\mathbb{I}_{kl}= \sigma^{z}_{kl}\rho \sigma^{z}_{kl} - \mathbb{I}_{kl}\rho \mathbb{I}_{kl}$, which is the familiar qubit dephasing expression projected onto the $\{|k\rangle,|l\rangle\}$ subspace, with $\sigma^{z}_{kl} = \sigma_{kk}-\sigma_{ll}= |k\rangle \langle k| - |l\rangle \langle l|$ and 
$\mathbb{I}_{kl} = \sigma_{kk}+\sigma_{ll}= |k\rangle \langle k| + |l\rangle \langle l|$.
%%%%%%%%%%

\subsection*{Experimental parameters and sample specifications}

For the $N=1$ and $N=2$ cases, experiments have been performed on a sample with $|0\rangle-|1\rangle$ and $|1\rangle-|2\rangle$ transition frequencies $\omega_{01}/(2\pi) = 5.01$ GHz and $\omega_{12}/(2\pi) = 4.65$ GHz. The simulations make use of the general form of the Lindblad	master equation for the quantum state evolution with relaxation and dephasing rates obtained from standard characterization measurements: $\Gamma_{10}=0.72$ MHz, $\Gamma_{21}=1.55$ MHz, $\Gamma_{10}^{\phi}=0.4$ MHz, $\Gamma_{21}^{\phi}=0.6$ MHz, and $\Gamma_{02}^{\phi}=1$ MHz.
 The duration of the beam-splitter pulse is $56$ ns (see also Eq.~\ref{eq:omegat}) and the amplitude of the pulse is directly proportional to the angle of rotation (in a given subspace). The $B$-pulses however have a fixed duration of $56$ ns until $\theta=3.38\pi$, beyond which the upper limit of the output power from our arbitrary waveform generator (AWG) is reached. To tackle this issue, the pulse duration is gradually increased from $56$ ns to $61$ ns in steps of $1$ ns (as 
$\theta$ varies from $3.38\pi$ to $4\pi$), such that the desired pulse-area is attained with lower pulse amplitudes. The transmon starts in thermal equilibrium at an effective temperature of $50$ mK (measured independently, see \cite{Sultanov2021}) such that the initial probability of occupation of the ground state, first excited state and 
second excited state is $p_0=0.9917 =99.17\%$, $p_1=0.0082=0.82\%$, and $p_2=0.0001=0.1\%$.

For experiments involving a large number of pulses ($N>2$) we use a sample with $\omega_{01}/(2\pi ) = 7.20$ GHz and $\omega_{12}/(2\pi) =6.85$ GHz. The relaxation and dephasing rates obtained from independent measurements are $\Gamma_{10}= 0.29$ MHz, $\Gamma_{21}=1.15$ MHz, $\Gamma_{10}^{\phi}=0.18$ MHz, $\Gamma_{21}^{\phi}=1.82$ MHz, and $\Gamma_{02}^{\phi}=1.70$ MHz.
 All the beam-splitter pulses are $56$ ns and $B$-pulses are of duration $112$ ns with various different amplitudes. For the case of identical $B$-pulses, $\theta$ is increased linearly from $0$ to $\pi$ in $180$ steps and in each case $p_0$ is measured for $N \in [1,25]$. To obtain the error bars, each experiment is repeated four times. In the case of random $B$-pulses, random strengths are chosen arbitrarily from a uniform distribution of random numbers from $0$ to $\pi$. Error bars result from the four repetitions of the same experiment. The corresponding surface maps, histograms and mean and standard deviation values are presented and discussed in the main text.
For further details on the errors due to pulse imperfections, see Supplementary Note 5~\cite{supp}.

For very long experiments, it is known that we can accumulate errors resulting in excess populations on the higher energy levels. The standard description for this effect is via an additional depolarizing channel~\cite{nielsen-book-2002}.
For a three-level system the depolarizing channel can be written in the operator-sum representation~\cite{kraus-book-1983}, which is a completely positive trace-preserving map, such that the final state is given by
%%%%%%%%%%%
\begin{equation}
 \rho_f = \sum_{\nu}K_{\nu} \rho K_{\nu}^{\dagger}, \quad \textrm{with} \quad
\sum_{\nu}K_{\nu}^{\dagger}K_{\nu}=\mathbb{I}_{3}. \label{Kraus}
 \end{equation}
 %%%%%%%%%%%%%%%
 The Kraus operators $K_{\nu}$'s are given in terms of
 Gell-Mann matrices: $K_1=\sqrt{\epsilon/6}\lambda_1$, 
 $K_2=\sqrt{\epsilon/6}\lambda_2$, $K_3=\sqrt{\epsilon/6}\lambda_4$, 
 $K_4=\sqrt{\epsilon/6}\lambda_5$, $K_5=\sqrt{\epsilon/6}\lambda_6$, 
 $K_6=\sqrt{\epsilon/6}\lambda_7$, $K_7=\sqrt{\epsilon}/3\lambda_3$,
 $K_8=\sqrt{\epsilon}/6(\sqrt{3}\lambda_8-\lambda_3)$,
 $K_9=\sqrt{\epsilon}/6(\sqrt{3}\lambda_8+\lambda_3)$, and 
 $K_{10}=\sqrt{1-8\epsilon/9}\,\mathbb{I}_{3}$. Here,
 $\lambda_{1(2)}=\sigma_{01}^{x(y)}$, $\lambda_{4(5)}=\sigma_{02}^{x(y)}$,
 $\lambda_{6(7)}=\sigma_{12}^{x(y)}$, $\lambda_{3}=\sigma_{01}^{z}$,
 and $\lambda_{8}=(\sigma_{02}^{z} + \sigma_{12}^{z})/\sqrt{3}$.
The final state following Eq.~(\ref{Kraus}) is 
\begin{equation}
	\rho_{f}= \frac{\epsilon \mathbb{I}_{3}}{3} + (1-\epsilon)\rho\;.
	\label{depol}
\end{equation}
In other words the system is replaced with the completely mixed state $\mathbb{I}_{3}/3$ with probability $\epsilon$ -- otherwise it is unaffected, with probability $1-\epsilon$. We consider only the depolarization caused by the $B$-pulse, with a value $\epsilon=1.8\times10^{-3}$ for a $\pi$ pulse applied on the $|1\rangle-|2\rangle$ transition; this is obtained by a best-fit of the  $\theta = \pi$ data. For arbitrary $\theta$ it is natural to consider a linear interpolation $\epsilon [\theta ]=1.8\times10^{-3}\times \theta/\pi$.

%%%%%%%%%%%%%%%%%%%%%%%%%%%%%%%%%%%%%%%%%
 \section*{Data availability}
Experimental and simulated data generated during this study are included in this published article (and its supplementary information files).
The experimental data that support the findings of this study can also be found in the GitHub repository \cite{github}.

 \section*{Code availability}
The codes for simulations that support the findings of this study can be found in the GitHub repository \cite{github}.

%%%%%%%%%%%%%%	


  %%%%%%%%%%%%%%%%%%%%%%%%%%%%%%%%%%%%%%%%%%%%%
  \section*{Acknowledgments}
  We are grateful to Kirill Petrovnin, Aidar Sultanov, Andrey Lebedev, Sergey Danilin, and Miika Haataja for assistance with sample fabrication and measurements. This project has received funding from the European Union's Horizon 2020 research and innovation programme under grant agreement no. 862644 (FET-Open project QUARTET). We also acknowledge support from the Academy of Finland under the RADDESS programme (project 328193) and the Finnish Center of Excellence in Quantum Technology QTF (projects 312296, 336810), as well as from Business Finland QuTI (decision 41419/31/2020). This work used the experimental facilities of the Low Temperature Laboratory and Micronova of OtaNano research infrastructure.

\section*{Author Contributions}
SD and GSP conceived the idea and obtained the key results.
SD performed the experiments and did a detailed analysis of 
the experimental data with inputs from JJM. SD and JJM did 
the numerical simulations. GSP supervised the project. All authors contributed to analytical calculations, discussed the results, and wrote the manuscript.

\section*{COMPETING INTERESTS}
The authors declare no competing interests.

  %%%%%%%%%%%%%%%%%%%%%%%%%%%%%%%%%%
  \newpage
  \begin{center} \textbf{\large{SUPPLEMENTARY INFORMATION}} \end{center}

  This supplement more thoroughly explores the coherent interaction-free measurements presented in this work, and discusses how our coherent interaction-free detection scheme compares with the standard projective non-unitary case typically realized in quantum optical systems \cite{Kwiat}. In particular, we compare efficiencies for both schemes up to $N = 25$, and further compare these two cases when dissipation is applied via the Lindblad master equation. We also present general analytical expressions for arbitrary $N$ for our coherent protocol and various simulations in support of our claims. An alternative analysis of the coherent interaction-free detection protocol is developed by considering the quantization of the $B$-pulse, which bears the same results as the ones obtained from the semi-classical description. This exercise helps contribute to an in-depth understanding of the process. Finally, we provide a geometric representation of the detection process on the Majorana sphere. We begin by presenting detailed analysis of interaction-free measurements in general and its coherent counterpart.

  \section*{Supplementary Note 1: Figures of merit}

  We introduce the key figures of merit for the $N=1$ case. For $N >1$, they can be generalized in straightforward ways. 
  The Positive Ratio $\mathrm{PR}=p_0(\theta)/(p_0(\theta)+p_1(\theta))$
  is a measure of the correct detection of a $B$-pulse with arbitrary strength ($\theta$) and the Negative Ratio $\mathrm{NR}=p_1(\theta)/(p_0(\theta)+p_1(\theta))$ is the incorrect non-detection of a $B$-pulse when it is applied with strength ($\theta$). Special cases are defined as follows: FPR and TNR correspond to $\theta=0$, while TPR  and FNR correspond to $\theta=\pi$ for PR and NR respectively. In fact, PR($\theta$) effectively corresponds to the number of instances that report an interaction-free measurement of the $B$-pulse and NR($\theta$) are the inconclusive outcomes, where both of these quantities are obtained by excluding the 
  situations where $B$-pulses are absorbed. In other words, for $N=1$ and $\theta=\pi$ we have a 50\% chance that the pulse is not absorbed. By postselecting over these cases, we  find that we can either sucessfully detect the pulse (with 50\% probability), or we cannot conclude anything (again with 50\% probability). This is the meaning of TPR and FNR, see also Fig. 2 in the main text.
  The extension of this logic for the case of large $N$ suggests that larger values of PR($\theta$) have direct correspondence with increasing probability of interaction-free/absorption-free detection of the $B$-pulse of strength $\theta$.
  
  \begin{table}[h]
  	\centering
  	\begin{tabular}{l|l|ll}
  		& \multicolumn{1}{c|}{Predicted positive}                                                                                                                                                                                                                                      & \multicolumn{1}{c}{Predicted negative}                                                                                                                                                                                                                                    &   \\
  		\cline{1-3}
  		\begin{tabular}[c]{@{}l@{}}\\Actual\\positive\\ \\ \end{tabular}  & \begin{tabular}[c]{@{}l@{}}\\$\rm TPR = \frac{p_{0}(\theta = \pi)}{p_{0}(\theta = \pi) + p_{1}(\theta = \pi)}$
  			\\ \\ \multicolumn{1}{c}{(True Positive Ratio)}\end{tabular} & 
  		\begin{tabular}[c]{@{}l@{}}\\$\rm FNR = \frac{p_{1}(\theta = \pi)}{p_{0}(\theta = \pi) + p_{1}(\theta = \pi)}$   
  			\\ \\ \multicolumn{1}{c}{(False Negative Ratio)}\end{tabular} \\
  		\cline{1-3}
  		\begin{tabular}[c]{@{}l@{}}\\Actual~\\negative\\ \\ \end{tabular} & \begin{tabular}[c]{@{}l@{}}\\$\rm FPR = \frac{p_{0}(\theta = 0)}{p_{0}(\theta = 0) + p_{1}(\theta=0)}$\\ \\ \multicolumn{1}{c}{(False Positive Ratio)}\end{tabular}   &                                                     \begin{tabular}[c]{@{}l@{}}\\$\rm TNR = \frac{p_{1}(\theta = 0)}{p_{0}(\theta = 0) + p_{1}(\theta = 0)}$    
  			\\ \\ \multicolumn{1}{c}{(True Negative Ratio)}\end{tabular}  
  		&   \\
  		\multicolumn{1}{l}{}                                      & \multicolumn{1}{l}{}                                                                                                                                                                                                                                                   &                                                                                                                                                                                                         &  
  	\end{tabular}
  	\caption{The confusion matrix as defined for the coherent detection of $\pi$ pulses. The concept is in fact applicable for the projective protocol as well, with the use of $p_{\mathrm{det}}$ and $1 - p_{\mathrm{det}} - p_{\mathrm{abs}}$ in place of $p_{0}$ and $p_{1}$ respectively.}
  	\label{predicted_ratios}
  \end{table}
  
  Borrowing from the standard terminology of hypothesis testing, we can introduce the confusion matrix for our detection protocol. We indicate the presence ($\theta = \pi$) or absence ($\theta = 0$) of a $B$-pulse as positive or negative respectively. The elements of the confusion matrix are summarized in Supplementary Table.~\ref{predicted_ratios}. Specifically, a true positive (TP) is the correct detection of an applied $\pi$ pulse (actual positive event), while a false positive (FP) is the incorrect prediction of the pulse when it has not been in fact applied (actual negative event). Strictly speaking, the complementary predictions are inconclusive in our case. However, for conformity, we will use the standard terminology of negative prediction to designate them, namely false negative (FN) and true negative (TN) for the cases when there was and was not a pulse present, respectively.

  In analogy with the optical case \cite{Kwiat_1995, Kwiat_1999}, we can also introduce the coherent interaction-free efficiency $\eta_{c} (\theta )=p_{0}(\theta ) /[p_{0}/(\theta ) + p_{2}(\theta )]$ as the fraction of pulses detected in an interaction-free manner while discarding the inconclusive results.
  
  It is also important to emphasize the role of setting the Ramsey sequence such that, in the absence of the pulse, the final state is $|1 \rangle$ and not say some superposition of $|0\rangle$ and $|1\rangle$. This ensures that, when finding the system in the state $|0\rangle$, we know with 100\% certainty that the pulse was present; in other words, that FPR=0 and TNR=1 in the ideal case.

  \section*{Supplementary Note 2: Coherent versus projective interaction-free measurements}
  
  We discuss here the difference between the standard non-unitary (projective) interaction-free measurement and our approach. To make the connection clear, we start with the $N=1$ case, for which simple analytical results can be provided. 
  
  \subsection{$N=1$ case}

  From the definitions in the main text we have
  \begin{eqnarray}
  	S_{1} &=& \frac{1}{\sqrt{2}}\mathbb{I}_{01} - \frac{i}{\sqrt{2}} \sigma_{01}^{y} + |2\rangle\langle 2|, \label{Eq-S1}
  	\\
  	B(\theta ) &=& |0\rangle\langle 0| + \cos\frac{\theta}{2}\mathbb{I}_{12} - i \sin\frac{\theta}{2} \sigma_{12}^{y},
  	\label{Eq-B}
  \end{eqnarray}
  where $\mathbb{I}_{kl}= |k\rangle \langle k| +  |l\rangle \langle l|$ and $\sigma_{kl}^{y}= -i|k\rangle \langle l| +i  |l\rangle \langle k|$.
  
  The corresponding Mach-Zehnder interferometric setup for non-unitary interaction-free measurements is shown in Supplementary Fig. \ref{fig:POVM}. The final experimental results are the events (clicks) recorded by the detectors $D_{0,1,2}$ modeled as projection operators onto the corresponding states, $D_{0}=|0\rangle \langle 0|$, 
  $D_{1}=|1\rangle \langle 1|$, and $D_{2}=|2\rangle \langle 2|$. By introducing a beam-splitter with finite reflectivity in the upper branch of the interferometer, we generalize the typical optical setups to the situation where the detector $D_2$ clicks only for a fraction of events.
  
  \begin{figure}
  	\centering
  	\includegraphics[width=0.8\linewidth]{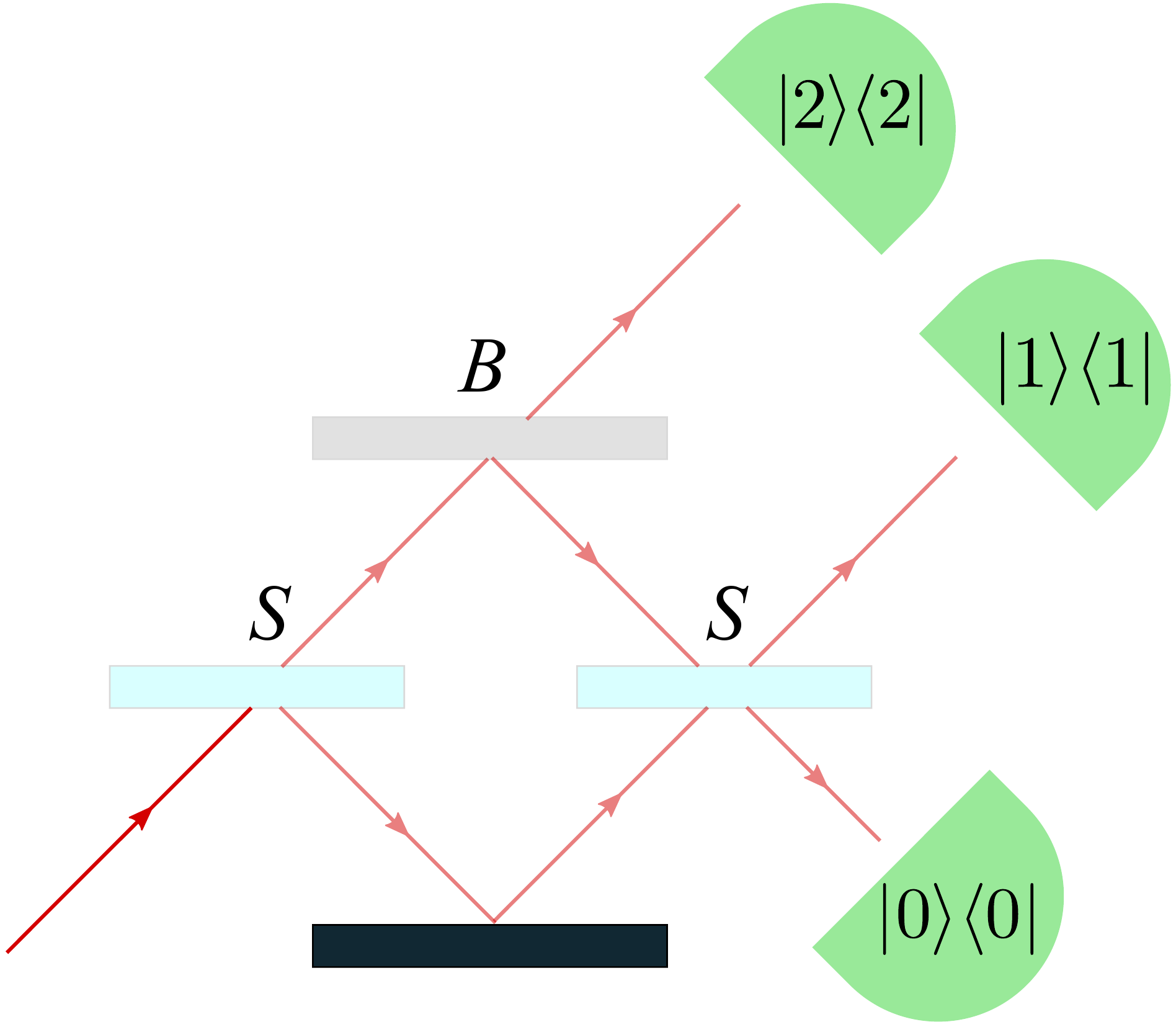}
  	\caption{Schematic of a generalized optical interaction-free interferometric setup, where the absorption probability can 
  		take values from 0\% to 100\%.}
  	\label{fig:POVM}
  \end{figure}
  
  To analyze what happens, let us first notice that in the absence of the $B$-pulse there is obviously no difference between the two approaches -- in both cases the evolution operator is the unitary $S_{1}^2 = -i \sigma_{01}^{y} + |2\rangle \langle 2|$. But, in the presence of a $B$-pulse, the superposition created by $S_1$ gets modified to 
  \begin{equation}
  	B(\theta )S_{1}|0\rangle = \frac{1}{\sqrt{2}}|0\rangle + \frac{1}{\sqrt{2}}\cos \frac{\theta}{2}|1\rangle +  \frac{1}{\sqrt{2}}\sin\frac{\theta}{2}|2\rangle .
  \end{equation}
  The measurement operators associated with the detector 
  $|2\rangle \langle 2|$ clicking (absorption) or not clicking (non-absorption) are the projectors
  \begin{eqnarray}
  	P_{\rm abs} &=& |2\rangle \langle 2|, \\
  	P_{\overline{\rm abs}} &=& |0\rangle \langle 0| + |1\rangle \langle 1|.
  \end{eqnarray} 
  Thus, with probability
  $p_{\rm abs}(\theta ) = \langle 0|S_{1}^{\dag}B^{\dag}(\theta )P_{\rm abs}B(\theta )S_{1}|0\rangle = (1/2) \sin^2 (\theta /2)$ the state of the system collapses to $\left(1/\sqrt{p_{\rm abs}(\theta )}\right)P_{\rm abs}B(\theta )S_{1}|0\rangle = |2\rangle$ if a click is recorded by the detector $|2\rangle \langle 2|$; otherwise,  with probability $p_{\overline{\rm abs}}(\theta ) = \langle 0|S_{1}^{\dag}B^{\dag}(\theta )P_{\overline{\rm abs}}B(\theta )S_{1}|0\rangle = 1/2 + (1/2) \cos^2 (\theta /2)$, the state collapses onto  $\left(1/\sqrt{p_{\overline{\rm abs}}(\theta )}\right)P_{\overline {\rm abs}}B(\theta )S_{1}|0\rangle =
  \left(1/\sqrt{1 + \cos (\theta /2)}\right)\left( |0\rangle + \cos (\theta /2) |1\rangle \right)$. Therefore, a non-absorption event has consequences: it confines the state to the $\{|0\rangle, |1\rangle \}$ manifold. For the case $\theta = \pi$, this confinement is onto the state $|0\rangle$ (we know for sure that the photon has traveled only in one branch of the interferometer), while the case $\theta =0$ corresponds to a completely reflective beam-splitter $B$, which fully hides the detector $|2\rangle \langle 2|$, and as a result the equal-weight superposition of $|0\rangle$ and $|1\rangle$ is not affected.  
  
  Note here that we can define the POVM measurement operators associated with the ensemble beam-splitter $B$ plus $|2\rangle \langle 2|$-detector from Supplementary Fig. \ref{fig:POVM} by $M_{\rm abs}=P_{\rm abs}B$ and $M_{\overline{\rm abs}}=P_{\overline{\rm abs}}B$, with the property $M_{\rm abs}^{\dag}M_{\rm abs} + M_{\overline {\rm abs}}^{\dag}M_{\overline{\rm abs}} = \mathbb{I}_{3}$, where $\mathbb{I}_{3}$ is the $3\times 3$  identity matrix. 
  
  The density matrix after the second beam splitter can be found by again applying $S_1$ to the states written above. Therefore, the state at the output is 
  \begin{equation}
  	\left(\sin^{2} \frac{\theta}{4}|0\rangle + \cos^{2} \frac{\theta}{4}|1\rangle\right)
  	\left(\sin^{2} \frac{\theta}{4}\langle 0| + \cos^{2} \frac{\theta}{4}\langle 1| \right)
  	+ \frac{1}{2} \sin^{2}\frac{\theta}{2}|2\rangle\langle 2|.\label{eq:out}
  \end{equation}
  
  As a result, the probability of interaction-free detection is $p_{\rm det}= \sin^{4}(\theta /4)$ (detector $|0\rangle\langle 0|$ clicks) and the efficiency $\eta$, defined as the fraction of successful detections by excluding the inconclusive cases ($|1\rangle\langle 1|$ clicks) is 
  \begin{equation}
  	\eta = \frac{p_{\rm det}}{p_{\rm det}+p_{\rm abs}} = \frac{2\sin^4(\theta/4)}{2\sin^4 (\theta/4)+
  		\sin^2(\theta/2)}. \label{eq:eta1}
  \end{equation}

  Consider now the coherent case. At the end of the protocol, the state is
  \begin{equation}
  	S_{1}B(\theta )S_{1}|0\rangle = \sin^{2}\frac{\theta}{4}|0\rangle + 	\cos^{2}\frac{\theta}{4}|1\rangle + \frac{1}{\sqrt{2}}\sin\frac{\theta}{2}|2\rangle .
  	\label{eq:finalN1}
  \end{equation}
  We can immediately verify that, by applying the same projectors $P_{\rm abs}$ and $P_{\overline{\rm abs}}$ corresponding to a measurement of the state $|2\rangle$, we obtain precisely the result Eq. (\ref{eq:out}).
  We have $p_0= \sin^{4}(\theta /4)$,  $p_0= \cos^{4}(\theta /4)$, $p_2 = (1/2)\sin^2 (\theta /2)$ and the coherent-case efficiency is
  \begin{equation}
  	\eta_{c} = \frac{p_{0}}{p_{0}+p_{2}} = \frac{2\sin^4(\theta/4)}{2\sin^4 (\theta/4)+
  		\sin^2(\theta/2)},
  \end{equation}
  the same as Eq. (\ref{eq:eta1}). This is due to the fact that $S_{1}P_{\rm abs}=P_{\rm abs}S_{1}$ and $S_{1}P_{\overline{\rm abs}}=P_{\overline{\rm abs}}S_{1}$, so it does not matter when we record the result of the projection on $|2\rangle \langle 2|$. 
  
  In conclusion, for $N=1$ there is no difference in the success/failure probabilities and the efficiency between the coherent and projective cases.
  The corresponding experimental results are shown in Supplementary Fig. \ref{fig:supplement-eta}, together with a comparison with the simulations and the ideal (decoherence-free) case.
  \begin{figure}[H]
  	\centering
  	\includegraphics[width=0.8\linewidth]{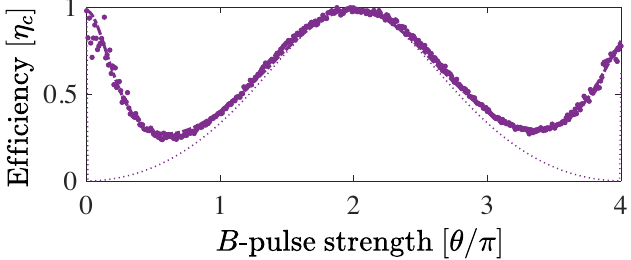} %{Main_fig_2.pdf}
  	\caption{
  		Corresponding to each $B$-pulse strength, the coherent efficiency $\eta_c (\theta )=p_0 (\theta )/[p_0 (\theta )+p_2 (\theta)]$ obtained from the experiment is shown as a purple line with circular markers, closely followed by the simulated curve (purple dot-dashed). The thin dotted purple line represents the respective ideal case, with no decoherence and without any experimental imperfections.}
  	\label{fig:supplement-eta}
  \end{figure}
  
  \subsubsection*{Quantizing the $B$-pulse in the $N=1$ case}
  
  We can get a deeper understanding of this effect by looking at the case where we treat the $B$-pulse quantum mechanically rather that in the semiclassical approximation. Let us denote by $b$ and $b^{\dag}$ the annihilation and creation operator describing the presence of photons from the $B$-pulse.
  In the rotating wave approximation, the interaction Hamiltonian between the pulse and the transmon is 
  \begin{equation}
  	H_{\rm int} = \frac{i\hbar g}{2} \left( b^{\dag} |1\rangle \langle 2|-
  	b|2\rangle \langle 1|\right). \label{eq:int}
  \end{equation}
  Consider now
  a Fock state $|\mathtt{n} \rangle$ with $\mathtt{n}$ photons. Experimentally, this can be realized as a cavity or resonator to which the transmon can be coupled and uncoupled.
  The Hamiltonian Eq. (\ref{eq:int}) conserves the number of excitations in the total Hilbert space of the resonator and the second transmon transition. As a result, the dynamics is confined to the subspace spanned by the vectors $|\mathtt{n}\rangle \otimes |1\rangle$ and $|\mathtt{n}-1\rangle \otimes |2\rangle$. In this subspace the Hamiltonian can be diagonalized; we obtain the eigenvectors
  \begin{eqnarray}
  	|\mathtt{n}+\rangle  &=& \frac{1}{\sqrt{2}} \left(  |\mathtt{n}\rangle \otimes |1\rangle  -i |\mathtt{n-1}\rangle \otimes |2\rangle
  	\right), \\
  	|\mathtt{n}-\rangle  &=& \frac{-1}{\sqrt{2}} \left(|\mathtt{n}\rangle \otimes |1\rangle + i|\mathtt{n}-1\rangle \otimes |2\rangle \right),
  \end{eqnarray}
  and the eigenvalues $E_{\pm} = \pm (\hbar /2) g \sqrt{\mathtt{n}}$,
  corresponding to a Rabi frequency $g\sqrt{n}$.
  Assume now a certain duration of the $B$-pulse --  let's denote it $\tau_{\rm B}$. We can define the corresponding strength $\theta_{\mathtt{n}}$ of the $\mathtt{n}$-photon pulse as $\theta_{\mathtt{n}} = g \sqrt{n}\tau_{\rm B}$. 
  We start in the state $|\mathtt{n}\rangle \otimes |0\rangle$ and apply $S_1$, $B$ (via the interaction Hamiltonian Eq.\ref{eq:int}), and again $S_{1}$. The final result is the state
  \begin{equation}
  	\sin^{2}\frac{\theta_{\mathtt{n}}}{4} |\mathtt{n}\rangle \otimes |0\rangle  + \cos^{2}\frac{\theta_{\mathtt{n}}}{4} |\mathtt{n}\rangle \otimes |1\rangle + \frac{1}{\sqrt{2}}\sin \frac{\theta_{\mathtt{n}}}{2} |\mathtt{n}-1\rangle \otimes |2\rangle .
  	\label{eq:Bquantum}
  \end{equation}
  The case when the cavity is not present ($g=0$) can be obtained directly from the expression above or by a separate calculation involving only the two consecutive $S_{1}$ unitaries, yielding, as expected, 
  \begin{equation}
  	|\mathtt{n}\rangle \otimes |1\rangle .
  \end{equation}
  First, we can immediately compare these results with the semiclassical expression Eq. (\ref{eq:finalN1}), to check that the probabilities are the same.
  But most importantly, Eq. (\ref{eq:Bquantum}) shows the entanglement and the energy balance between the pulse and the detector: if the transmon is found in the state $|0\rangle$ then the pulse will still contain $\mathtt{n}$ photons, \emph{i.e.}, no photon has been absorbed. On the contrary, if the transmon is found in state $|2\rangle$, this could happen only with the absorption of a photon from the $B$-pulse. In the case that state $|1\rangle$ is detected we cannot conclude anything, but we can still rest assured that the cavity is not affected even if it was present in the setup.

  \subsection{$N=2$ case}

  For the $N=2$ case the beam splitter $S_{2}$ is a $\pi/3$ pulse
  \begin{equation}
  	S_{2} = \frac{\sqrt{3}}{2} \mathbb{I}_{01} -  \frac{i}{2}\sigma_{01}^{y} +|2\rangle \langle 2|.  \label{Eq-S2}
  \end{equation}
  The final state is
  \begin{equation}
  	\begin{split}
  		\frac{1}{8}\Big{[}3\sqrt{3} -\sqrt{3}\cos \frac{\theta_{1}}{2} - 2\sqrt{3}\cos^{2} \frac{\theta_{1}}{4} \cos \frac{\theta_{2}}{2} \\+ 2\sin \frac{\theta_{1}}{2} \sin \frac{\theta_{2}}{2} \Big{]}|0\rangle + \frac{1}{8}\Big{[}3 - \cos \frac{\theta_{1}}{2}  \\ + 6\cos^{2} \frac{\theta_{1}}{4} \cos \frac{\theta_{2}}{2} - 2\sqrt{3}\sin \frac{\theta_{1}}{2} \sin \frac{\theta_{2}}{2} \Big{]}|1\rangle \\ + \frac{1}{2}\Big{[}\sin \frac{\theta_{1}}{2} \cos \frac{\theta_{2}}{2} + \sqrt{3} \cos^{2} \frac{\theta_{1}}{4} \sin \frac{\theta_{2}}{2}\Big{]}|2\rangle\;. \label{eq:N2semiclassical}
  	\end{split}
  \end{equation} 
  At maximum strength $\theta_{1}=\theta_{2} =\pi$ this state reads
  \begin{equation}
  	\frac{1}{8}(2+3\sqrt{3})|0\rangle + \frac{1}{8}(3-2\sqrt{3})|1\rangle + \frac{1}{4}\sqrt{3}|2\rangle .
  \end{equation}
  We can already see that the probability of absorption is $p_2=3/16 = 0.1875$, smaller than the 0.25 of the single-interrogation detection, and the probability $p_0$ of an IFM detection is $p_0=(31+12\sqrt{3})/64 \approx 0.8091$, significantly larger than the $0.25$ of the single-interrogation case. The efficiency of the coherent detection is 
  \begin{equation}
  	\eta_{c} = \frac{p_0}{p_0+p_2} = 0.8118.
  \end{equation}
  Note that the efficiency is so high because the probability of failing to find the pulse is very small, $p_1 = 0.0034$. 
  
  Further, we experimentally realize a general 
  case where $B$-pulse strengths are different, \emph{i.e.},
  $\theta_1, \theta_2 \in [0, 4\pi]$. Maps of the experimental and simulated 
  results for the efficiency $\eta_c$ are 
  shown as functions of $\theta_1$ and $\theta_2$ in Supplementary Fig.~\ref{fig:012}. 
  The variation of the ground state, first excited state and second excited 
  state probabilities as functions of $\theta_1$ and $\theta_2$ is shown in
  the main text alongside with the Positive Ratio $PR(\theta_1, \theta_2)=p_0(\theta_1, \theta_2)/(p_0(\theta_1, \theta_2)+p_1(\theta_1, \theta_2))$ and the Negative Ratio $NR(\theta_1, \theta_2)=p_1(\theta_1, \theta_2)/(p_0(\theta_1, \theta_2)+p_1(\theta_1, \theta_2))$.
  Experimental and simulated results are in very good agreement with each other.
  % as a function of $\theta_1$ and $\theta_2$ are shown in Fig.\ref{fig:012PRNR}.

  \begin{figure}[h]
  	\centering
  	\includegraphics[width=0.87\linewidth]{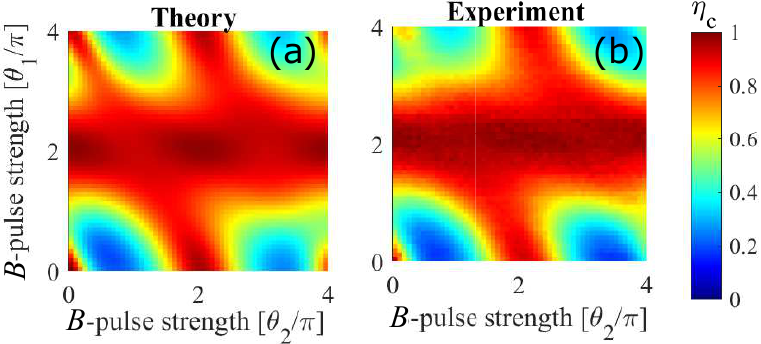}
  	\caption{Simulated and experimental map of coherent efficiency $\eta_c$ as a function of $\theta_1$ and $\theta_2$.} 
  	\label{fig:012}
  \end{figure}

  % \begin{figure}[h]
  	% 	\centering
  	% 	\includegraphics[width=0.9\linewidth]{MZI_2b2d_PR_NR2.jpg}
  	% 	\caption{(a,c) Simulated and (b,d) Experimental 2D maps for the (a,b) Positive Ratio $\mathrm{PR}(\theta_1, \theta_2)$ and the (c,d) Negative Ratio $\mathrm{NR}(\theta_1, \theta_2)$ as a function of $\theta_1$ and $\theta_2$. }
  	% 	\label{fig:012PRNR}
  	% \end{figure}

  Importantly, we also observe that the previous maximum at $\theta =2 \pi$ from the $N=1$ case (see Fig. 2 in the main text) starts to flatten, evolving towards  becoming a plateau, a tendency that will become even more prominent for $N \gg 1$.

  Let us now clarify the difference with respect to the standard projective (non-unitary) interaction-free measurement, considering for simplicity the case $\theta_{1}=\theta_{2}=\pi$.
  After the first pulse of strength $\theta_{1}$ the state becomes
  \begin{equation}
  	B(\theta_{1}) S_{2} |0\rangle = \frac{\sqrt{3}}{2}|0\rangle + \frac{1}{2} \cos\frac{\theta_{1}}{2} |1\rangle + \frac{1}{2} \sin\frac{\theta_{1}}{2} |2\rangle .\label{eq:sm}
  \end{equation}
  This is the state that serves as the input for the next Ramsey $S_2$ pulse. 
  
  We can now see that there is a crucial difference with respect to the case when there has been a measurement of the second excited state and the result was negative. In this situation, the state entering the second $S_2$ pulse is 
  \begin{eqnarray}
  	\frac{1}{\sqrt{p_{\overline{\rm abs}}(\theta_1)}}P_{\overline{\rm abs}}B(\theta_{1})S_{2}|0\rangle = \nonumber \\ =\frac{1}{\sqrt{3+\cos^{2}(\theta_{1}/2)}}\left(\sqrt{3}|0\rangle + \cos\frac{\theta_{1}}{2}|1\rangle \right)  \label{eq:bm}
  \end{eqnarray}
  where $p_{\overline{\rm abs}}(\theta_1)=\frac{3}{4} + \frac{1}{4} \cos^2 (\theta_{1}/2)$. Unlike Eq. (\ref{eq:sm}), this state does not have a component on $|2\rangle$. In the case $\theta_1 = \pi$, the state Eq. (\ref{eq:bm}) seen by the second $S_2$ pulse is $|0\rangle$, the same as the initial one. Thus, the same interference phenomena is reproduced in the second Ramsey cycle.  In contrast, for the coherent case, Eq. (\ref{eq:sm}) contains a component on $|2\rangle$, which precisely encapsulates our lack of knowledge about the $p_2$ probability at the beginning of the second Ramsey cycle. 
  
  We can also calculate the probabilities and efficiency in the non-unitary case for $\theta_{1}=\theta_{2}=\pi$,  $p_{\rm det}= (\cos^2(\pi/6))^3 = 27/64 = 0.4219$ and $p_{\rm abs}=\sin^{2}(\pi/6)[1 + \cos^{2}(\pi/6)]=7/16 = 0.4375$, resulting in an efficiency
  \begin{equation}
  	\eta= \frac{p_{\rm det}}{p_{\rm det}+p_{\rm abs}}=0.49 .
  \end{equation} 
  We see for the case of $N=2$ that the efficiency of the coherent case is significantly larger!

  \subsubsection*{Quantizing the $B$-pulse in the $N=2$ case}
  
  In a similar way to the $N=1$ case, we can treat the $B$-pulse quantum mechanically. We consider that an interaction Hamiltonian Eq. (\ref{eq:int}) is available, such that the transmon can be coupled in a controllable way to the field.
  
  Suppose that the transmon is coupled in both sequences to the same mode containing $\mathtt{n}$ photons. These photons can be for example located in a cavity, which is coupled by a tunable coupling element to the transmon, or they can be traveling in a transmission line, as in our experiments. The initial state is 
  $|\mathtt{n}\rangle \otimes |0\rangle$.
  The final state can be obtained by the same procedure as in the $N=1$ case, and reads
  \begin{eqnarray}
  	&&		\frac{1}{8} \left[ 3\sqrt{3} - \sqrt{3}\cos \frac{\theta_{1\mathtt{n}}}{2}
  	- 2\sqrt{3}\cos^{2} \frac{\theta_{1\mathtt{n}}}{4}\cos \frac{\theta_{2\mathtt{n}}}{2} + \right. \nonumber\\
  	&&		\left.
  	+2\sin \frac{\theta_{1\mathtt{n}}}{2} \sin \frac{\theta_{2\mathtt{n}}}{2} \right]|\mathtt{n}\rangle \otimes |0\rangle + \nonumber\\
  	&&		\frac{1}{8}\left[ 3 - \cos \frac{\theta_{1\mathtt{n}}}{2}  
  	+ 6\cos^{2} \frac{\theta_{1\mathtt{n}}}{4} \cos \frac{\theta_{2\mathtt{n}}}{2} \right. \nonumber \\
  	&& 		\left. - 2\sqrt{3}\sin \frac{\theta_{1\mathtt{n}}}{2} \sin \frac{\theta_{2\mathtt{n}}}{2} \right] |\mathtt{n} \rangle \otimes
  	|1\rangle + \nonumber \\
  	&&		+\frac{1}{2}\left[
  	\sin \frac{\theta_{1\mathtt{n}}}{2} \cos \frac{\theta_{2\mathtt{n}}}{2} + \sqrt{3} \cos^{2} \frac{\theta_{1\mathtt{n}}}{4} \sin \frac{\theta_{2\mathtt{n}}}{2}\right]|\mathtt{n}-1 \rangle \otimes |2\rangle \nonumber \\
  \end{eqnarray} 
  with the notation $\theta_{1\mathtt{n}} = g\sqrt{n}t_{\rm B1}$ and $\theta_{2\mathtt{n}} = g\sqrt{n}t_{\rm B2}$.
  We immediately observe the similarity with the semiclassical result Eq. (\ref{eq:N2semiclassical}). The result very clearly reaffirms that the photonic Fock state does not change by finding the qubit in the state $|0\rangle$. It can lose a photon only if the level $|2\rangle$ is excited. Thus, we can detect the existence of photons inside the cavity without absorbing any of them.
  
  {\it Generalization to two different modes:} We can also imagine the situation when the  transmon is coupled to different modes in the two sequences, for example realized as photons in two distinct cavities. Suppose that in the first sequence it interacts with a cavity containing $\mathtt{n}$ photons, while in the second sequence it interacts with another cavity, containing $\mathtt{n}$ photons. 
  The initial state is then $|\mathtt{m},\mathtt{n}\rangle \otimes |0\rangle$. 
  The final state in this case can be calculated as
  \begin{eqnarray}
  	& &		\frac{1}{8} \left[\left(3 \sqrt{3} - \sqrt{3} \cos \frac{\theta_{1\mathtt{n}}}{2}
  	- 2\sqrt{3}\cos^{2} \frac{\theta_{1\mathtt{n}}}{4}\cos \frac{\theta_{2\mathtt{m}}}{2}\right) |\mathtt{m},\mathtt{n}\rangle \right. \nonumber \\
  	&&	    \left.	+ 2\sin \frac{\theta_{1\mathtt{n}}}{2} \sin \frac{\theta_{2\mathtt{m}+1}}{2}|\mathtt{m+1},\mathtt{n-1} \rangle \right] \otimes |0\rangle + \nonumber \\
  	&&		\frac{1}{8}\left[\left( 3 - \cos \frac{\theta_{1\mathtt{n}}}{2} 
  	+ 6\cos^{2} \frac{\theta_{1\mathtt{n}}}{4} \cos \frac{\theta_{2\mathtt{m}}}{2}\right) |\mathtt{m},\mathtt{n}\rangle \right. \nonumber \\
  	&&    \left.  - 2\sqrt{3}\sin \frac{\theta_{1\mathtt{n}}}{2} \sin \frac{\theta_{2\mathtt{m}+1}}{2}|\mathtt{m+1},\mathtt{n-1}\rangle \right]
  	\otimes |1\rangle \nonumber \\
  	&&	 + \frac{1}{2}\left[
  	\sin \frac{\theta_{1\mathtt{n}}}{2} \cos \frac{\theta_{2\mathtt{m}+1}}{2} |\mathtt{m},\mathtt{n-1}\rangle + \right. \nonumber \\				
  	&& \left. \sqrt{3} \cos^{2} \frac{\theta_{1\mathtt{n}}}{4} \sin \frac{\theta_{2\mathtt{m}}}{2}|\mathtt{m-1},\mathtt{n}\rangle \right]\otimes |2\rangle ,
  \end{eqnarray} 
  with the notation $\theta_{1\mathtt{n}} = g_{1}\sqrt{n}t_{\rm B1}$ and $\theta_{2\mathtt{m}} = g_{2}\sqrt{m}t_{\rm B2}$,  $\theta_{2\mathtt{m}+1} = g_{2}\sqrt{m+1}t_{\rm B2}$.
  If the transmon gets excited, we see that this can happen with the loss of a photon from either one of the modes.
  If the transmon is found in the state $|0\rangle$, then we can ascertain the existence of photons in the cavities, and, at the same time, we have transformed the initial Fock state $|\mathtt{m},\mathtt{n}\rangle$ into a coherent superposition of $|\mathtt{m},\mathtt{n}\rangle$ and $|\mathtt{m+1},\mathtt{n-1}\rangle$. The latter of course represents the possibility that a photon gets absorbed by the transmon during the first Ramsey sequence and reemitted into the same cavity during the second sequence.
  The transformation of a Fock state into a coherent state is a feature that
  is reminiscent of the famous Hanbury Brown-Twiss experiment \cite{Brown_1956}. 
  
  \subsection{$N > 1$ case}
  
  We have seen that for $N=1$ the efficiency of the coherent protocol is the same as that of the projective protocol, while for $N=2$ the coherent protocol is more advantageous. Does this tendency continues for large $N$? Let us take one more step and look at the case $N=3$. In the coherent protocol, the output state is 
  \begin{eqnarray}
  	& & \cos\frac{\pi}{8}\left(\cos^{3}\frac{\pi}{8} + 2 \sin^2\frac{\pi}{8}\right)
  	|0\rangle  \nonumber
  	\\
  	& & + \sin\frac{\pi}{8}\left(\cos^{3}\frac{\pi}{8} + \sin^2\frac{\pi}{8} - \cos^2\frac{\pi}{8}
  	\right)|1\rangle  \nonumber
  	\\
  	& & + \sin\frac{\pi}{8}\cos\frac{\pi}{8}\left(\cos\frac{\pi}{8} -1 \right)|2\rangle .
  \end{eqnarray} 
  We can verify immediately that $p_{0} > \cos^{8}(\pi/8) = p_{\rm det}$ and $p_{2} < \sin^{2}(\pi/8) (1 + \cos^2 (\pi /8) + \cos^4 (\pi /8)) = p_{\rm abs}$, where $p_{\rm det}$ and  $p_{\rm abs}$ are the detection and  absorption probabilities in the $N=3$ case, respectively.

  We can now generalize the protocol to $N$ $B$-pulses and the same number of Ramsey sequences. In this case the $S$ pulses are defined as
  \begin{equation}
  	S_{N} = \exp \left[-i\frac{\pi}{2(N+1)} \sigma^{y}_{01}\right].
  \end{equation}

  The efficiency of  the coherent detection is defined as before:
  \begin{equation}
  	\eta_{c} = \frac{p_0}{p_0+p_2}.
  \end{equation}

  Let us now consider the non-unitary (projective) protocol. In this case the probability of 
  a successful detection is the product of probabilities that the system stays in the state $|0\rangle$
  \begin{equation}
  	p_{\rm det} = \left[\cos\left(\frac{\pi}{2(N+1)}\right)\right]^{2(N+1)}, \label{eq:detproj}
  \end{equation}
  while the absorption probability is
  \begin{equation}
  	p_{\rm abs} = \sin^{2}\left(\frac{\pi}{2(N+1)}\right)\sum_{n=0}^{N-1} \left[\cos\left(\frac{\pi}{2(N+1)}\right)\right]^{2n}.
  \end{equation}
  Note that $\left[\cos\left(\pi/2(N+1)\right)\right]^{2(N+1)} \overset{N\gg 1}{\approx} 1 - \pi^2/4(N+1) + \mathcal{O}[N^{-2}]$, therefore $p_{\rm det}$ approaches unity at large N.
  The second expression is a sum of independent probabilities (that there is absorption in the first Ramsey sequence, that there is no absorption in the first Ramsey sequence but there is in the second, etc.). The efficiency is 
  \begin{equation}
  	\eta= \frac{p_{\rm det}}{p_{\rm det}+p_{\rm abs}}.
  \end{equation}
  
  The efficiencies obtained in the coherent and the
  projective cases for N$\in [1,25]$ are plotted in Supplementary Fig.~\ref{fig:efficiencies},
  with and without decoherence. Clearly, the efficiency obtained in the 
  coherent case  is significantly higher than that of the projective
  case, \emph{i.e.}  $\eta_c > \eta$ and  $\eta_c^{(d)} > \eta^{(d)}$ for any value of $N>1$.
  In the presence of decoherence, the difference between the two cases tends to stay constant with increasing $N$. 
  
  \begin{figure}[H]
  	\centering
  	\includegraphics[width=0.8\linewidth]{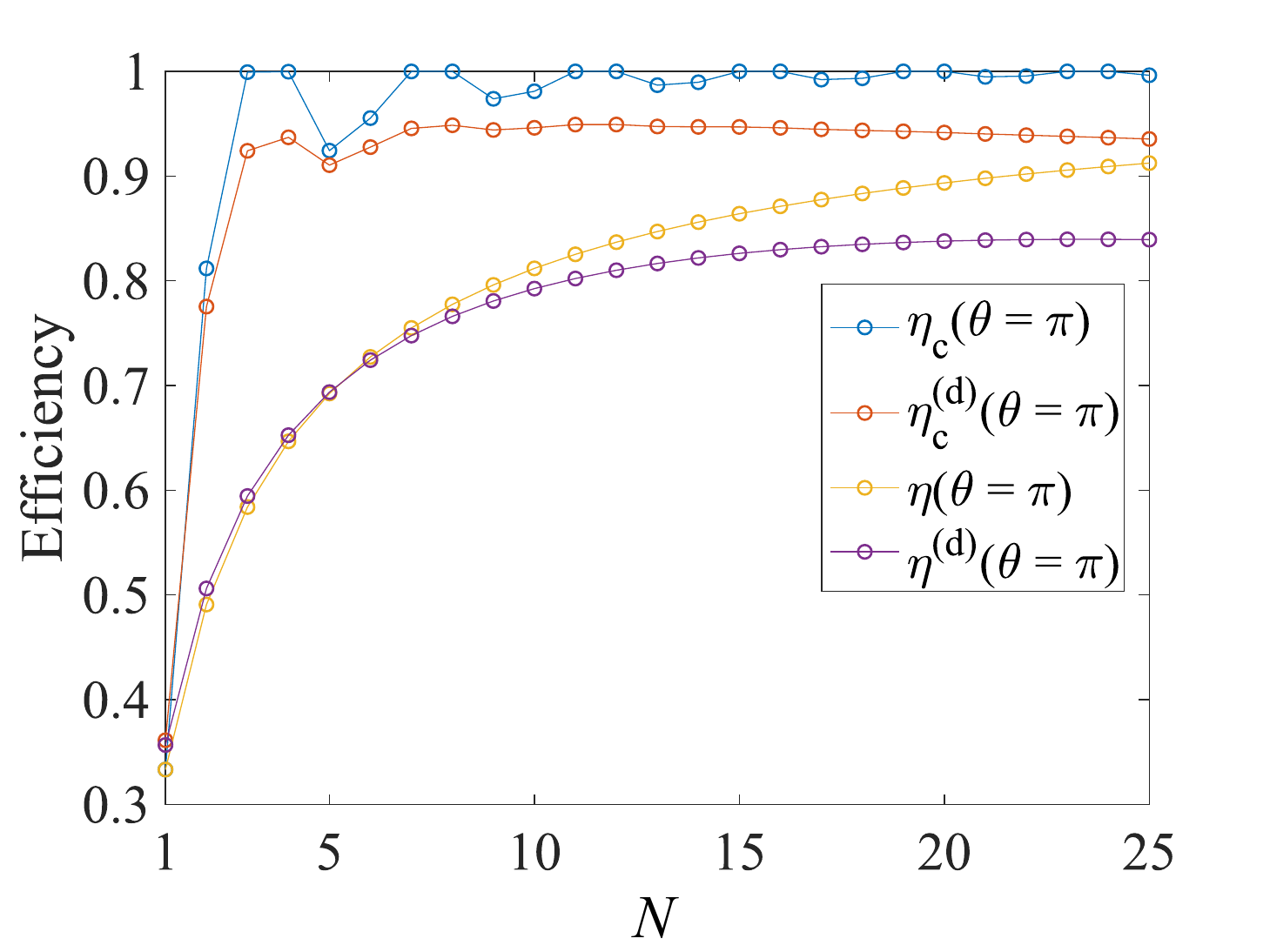}
  	\caption{The efficiency of our scheme $\eta_{c}$ as a function of $N$ Ramsey segments is compared with the efficiency of the standard projective scheme $\eta$, all at strength $\theta = \pi$. The corresponding cases with dissipation included are denoted by $\eta_{c}^{(d)}$ and $\eta^{(d)}$. 
  	}
  	\label{fig:efficiencies}
  \end{figure}
  
  {\it Elements of the confusion matrix in coherently repeated interrogations.}
  \begin{figure}[ht]
  	\centering
  	\includegraphics[width=8cm,keepaspectratio=true]{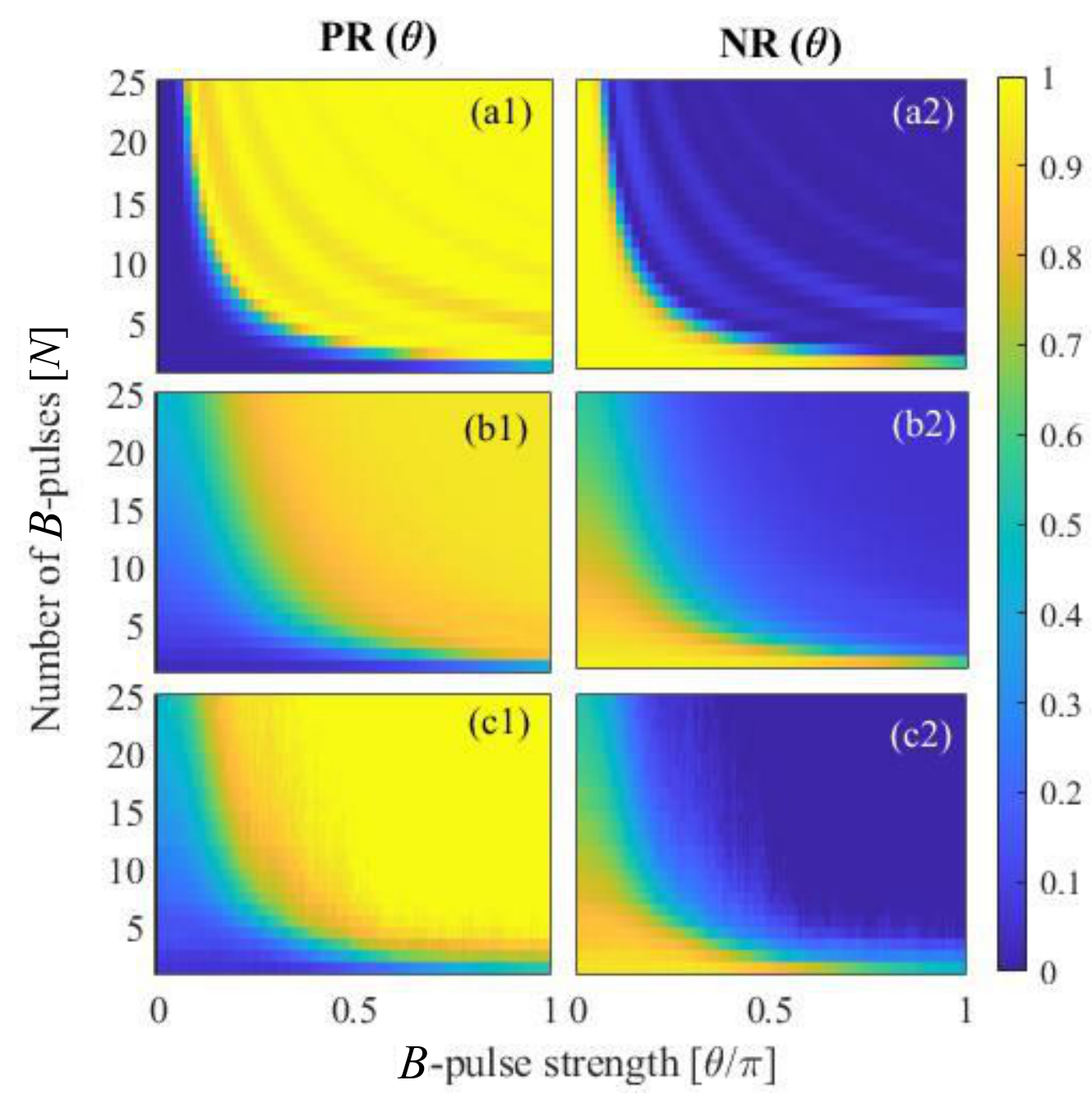}
  	%\includegraphics[width=8cm,keepaspectratio=true]{MZI_revision_fig1.jpg}
  	% MZI_revision_fig1.jpg: 510x511 px, 72dpi, 17.99x18.03 cm, bb=0 0 510 511
  	\caption{(a1, b1, c1) Positive ratio, PR ($\theta$) and (a2, b2, c2) Negative ratio, NR ($\theta$) are plotted for different number of $B$-pulses N with equal strengths $\theta$, where $\theta \in [0,\pi]$ linearly sweeps the range of $B$-pulse strengths for a given $N \in [1,25]$. Panels (a1,a2) present the ideal situation with no decoherence, panels
  		(b1,b2) are the results obtained from simulation with decoherence and panels (c1,c2) correspond to the results from the experiments.}
  	\label{fig:Nconfusion}
  \end{figure}
  We can obtain the elements of the confusion matrix in a more general form for the case of $N$ $B$-pulses, see Supplementary Fig.~(\ref{fig:Nconfusion}). 
  The general 2D maps of these 
  positive and negative ratios are plotted as functions of the number of $B$-pulses and $B$-pulse strength as shown in Supplementary Fig.~(\ref{fig:Nconfusion}) for ideal simulation without decoherence, simulation with decoherence, and results from the experiments. 
  It is clear from the surface maps in Supplementary Fig.~(\ref{fig:Nconfusion}) that the True Positive Ratio (TPR) is close to $1$ and the False Negative Ratio (FNR) is close to $0$ for $N>2$ as observed from the simulated and the experimental results. Ideally, FPR and TNR are independent of $N$, but there is an increase in FPR values and a decrease in TNR values with increasing $N$ in parts (b1,b2,c1,c2) of Supplementary Fig.~(\ref{fig:Nconfusion}), which is due to the long sequences, where decoherence is significant.
  The experimental data used in this section for arbitrary $N$ correspond to the case of equal $B$-pulse strengths varying linearly between $[0,\pi]$; consistent with the data shown in Fig. 6(a) of the main text.

  We also obtain the coherent interaction-free efficiency $\eta_c$ as a function of the $B$-pulse equal-strength $\theta$ and number of $B$-pulses $N$. The simulated values of $\eta_c$ are shown as a surface plot in Supplementary Fig.~\ref{fig-eta-2d}(a) with a few experimental values for various combinations of ($N, \theta$) marked on top of the surface plot. The continuous black curve corresponds to the simulated values of $\eta_c=0.85$. Supplementary Fig.~\ref{fig-eta-2d}(b) shows the simulated (continuous line) and experimental values (black circular markers) of $\eta_c$ at maximum $B$-pulse strength $\theta=\pi$ at various $N$'s. 
  Clearly, the simulation and corresponding experimental values depict a wide region of highly efficient interaction-free detection of the $B$-pulses.
  
  \begin{figure}[ht]
  	\centering
  	\includegraphics[width=8cm,keepaspectratio=true]{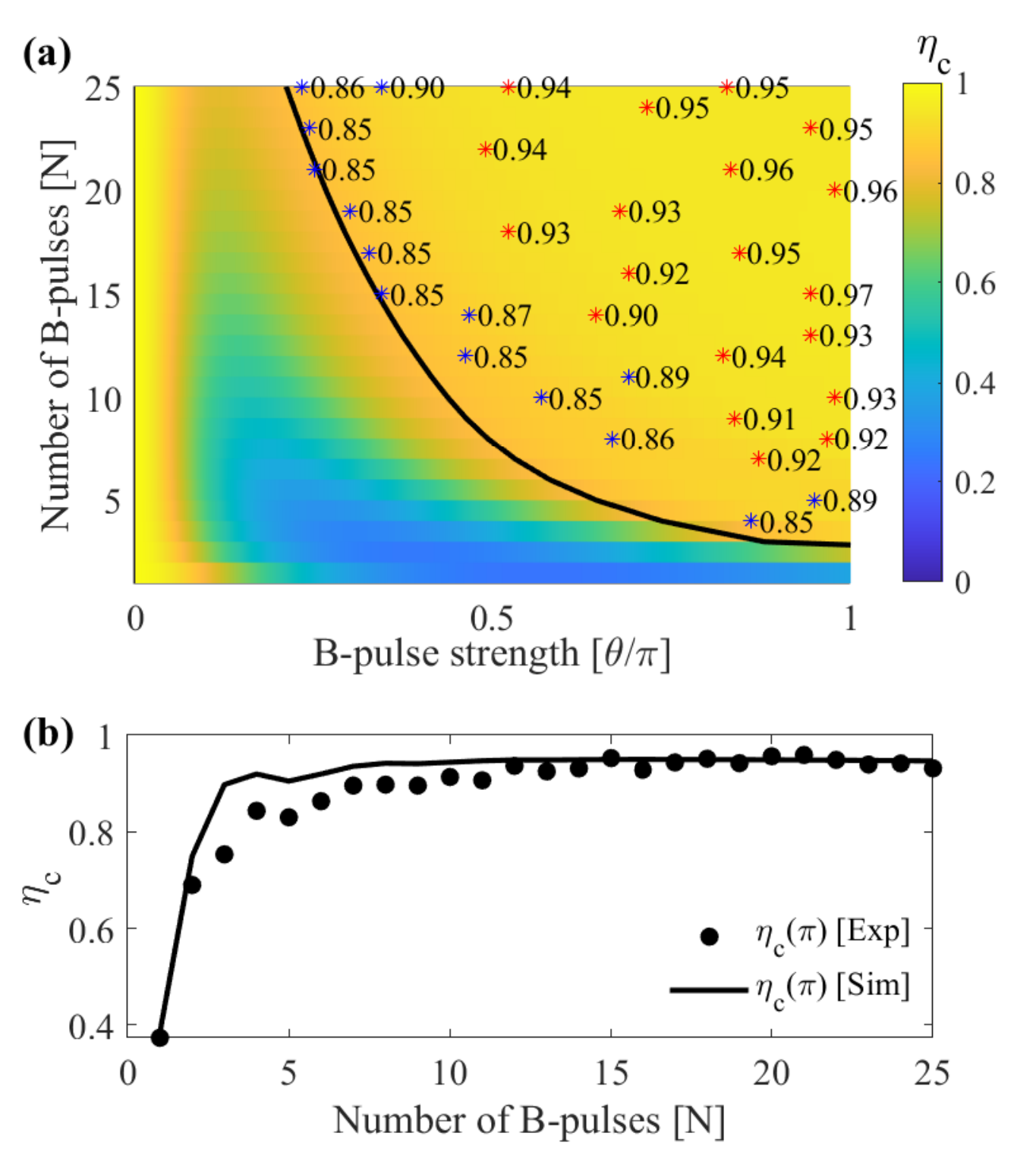}
  	% MZI_revision_fig1.jpg: 510x511 px, 72dpi, 17.99x18.03 cm, bb=0 0 510 511
  	\caption{(a) Surface plot of simulated $\eta_c$, plotted versus the number of $B$-pulses and $\theta$ with explicit experimental values marked on top of it. Blue markers correspond to experimental $\eta_c$ in the range [$0.85, 0.90$) and red markers correspond to experimental $\eta_c>0.90$. The continuous black curve corresponds to simulated $\eta_c=0.85$. The formation of a plateau of high efficiency values is thus confirmed experimentally.
  		(b) Coherent interaction-free efficiency is plotted as a function of $N$ for $\theta=\pi$, where the continuous back curve presents the simulated profile $\eta_c(\theta =\pi)$ and the circular markers correspond to the respective experimental values.}
  	\label{fig-eta-2d}
  \end{figure}
  
  Next, let us reconsider the $p_0$ profiles for various different values 
  of $N$  as a function of $\theta$. %, where for a given $N$, all the $B$-pulses are of the same strength $\theta$, varying linearly between $[0,\pi]$ as shown in Fig. 4(a) of the main text.
  As expected, $p_0$ gradually rises from $0$ to a maximum value with increasing $\theta$ and then tends to stay higher, forming a plateau which
  is symmetrical around $\theta=2\pi$. This plateau gets wider with increasing $N$.
  We quantify the widening in terms of the area $\int_{0}^{\pi} d \theta p_{0}(\theta )$ enclosed under $p_0$ -- as a function of $\theta$ --  for a given $N$
  and for $\theta\in[0, \pi]$. The results for equal and unequal arbitrary $B$-pulse strengths are shown in Supplementary Fig.~\ref{fig-plateau} (a,b) respectively. Evaluation of area 
  from the experimental data is shown with red square markers, with the simulation as continuous black curve. The dotted black curve is the simulation without considering the depolarization channel, while the dashed black curve signifies the ideal case without decoherence. As expected, the simulation in the absence of depolarization 
  predicts higher values than without depolarization, while the ideal case provides the upper limit to the area. Note also that the respective plots of area for unequal $B$-pulse strengths are higher than for equal $B$-pulse strengths. Once again this conveys the idea that unequal random $B$-pulse strengths give rise to higher efficiency of coherent interaction-free detection.
  
  \begin{figure}[h]
  	\centering
  	\includegraphics[width=0.8\linewidth]{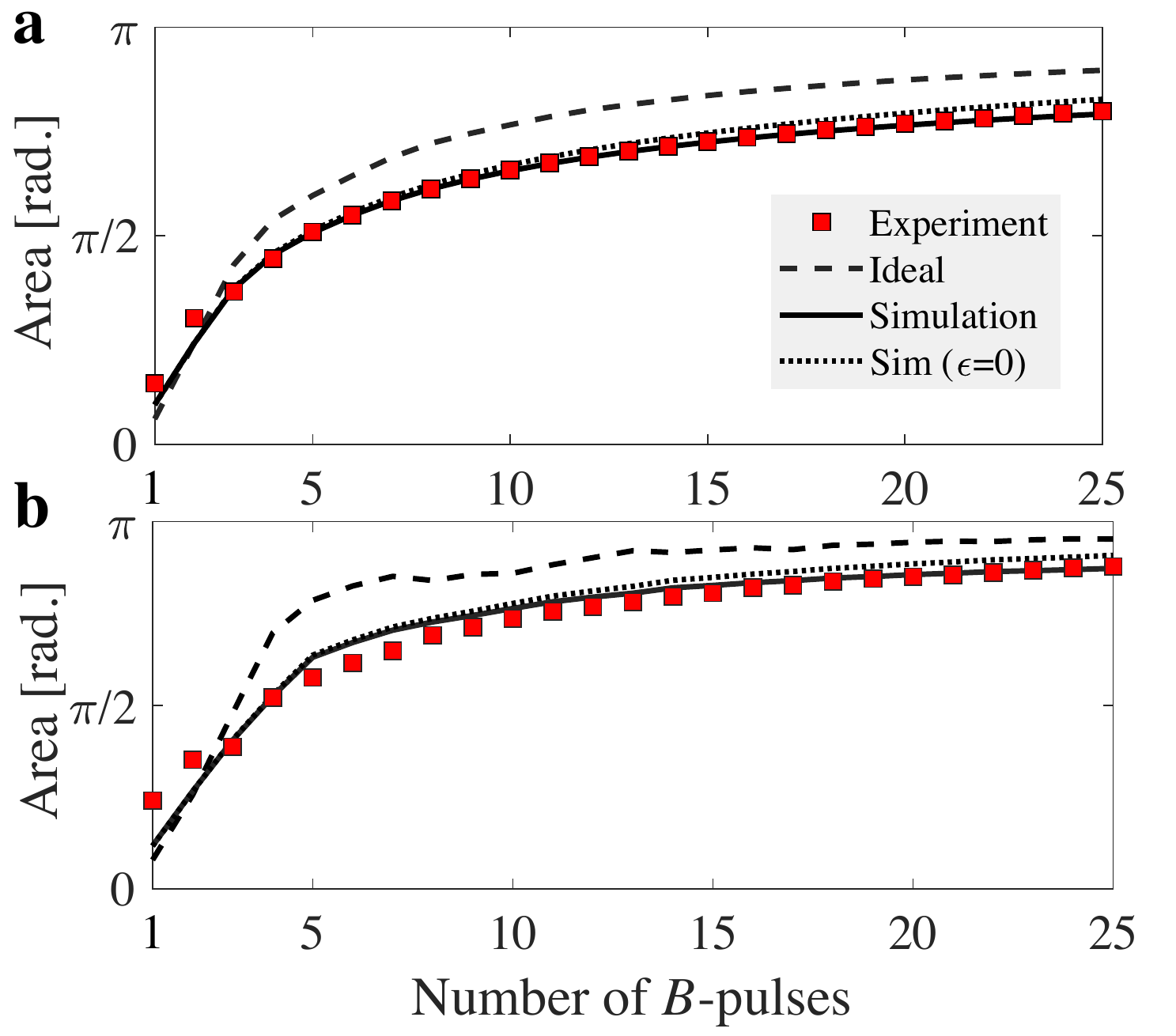}
  	\caption{Area under the 
  		$p_0 (\theta )$ plots as a function of $N$ with 
  		$\theta\in [0, \pi]$ 
  		for (a) Identical $B$-pulses with linearly varying strengths, (b) Randomly chosen $B$-pulses.}
  	\label{fig-plateau}
  \end{figure}
  %%%%%%%%%%% 
  
  Another important observation is that, in order to work properly, the protocol should start with the transmon in state $|0\rangle$. This is because the imbalance in the beam-splitter is designed such that the $B$-pulse is probed only weakly at each pass, with most of the weight of the superposition meant to stay in the $|0\rangle$ state. This, of course, is also the case for the optical projective realizations. To understand this better, we can simulate the situation where we start in state $|1\rangle$ for the case of uniform values $\theta = \pi$, see Supplementary Fig. \ref{fig-initialstates}. One can see that if the protocol is run correctly, with the ground state as the initial state, the probabilities stabilize relatively fast to the values $p_{0}\approx 1$, $p_{1}\approx p_{2} \approx 0$. But in the case when we start with $|1\rangle$, the $\theta =\pi$ excitation is shuffled between the transmon and the pulse and the protocol does not yield some stationary values. Indeed, the state after each odd pulse $N$ leaves the transmon in the state $|2\rangle$ so nothing happens at  the $N+1$ beam-splitter. Then, at the next encounter with the pulse (even $N+1$) the transmon goes in the state $|1\rangle$ by stimulated emission. As it encounter the $N+2$ beam-splitter, the transmon remains mostly on the state $|1\rangle$ due to the asymmetry of  the beam-splitter. Then it sees again an odd $N$+2 pulse, etc.
  
  \begin{figure}[h]
  	\centering
  	\includegraphics[width=0.9\linewidth]{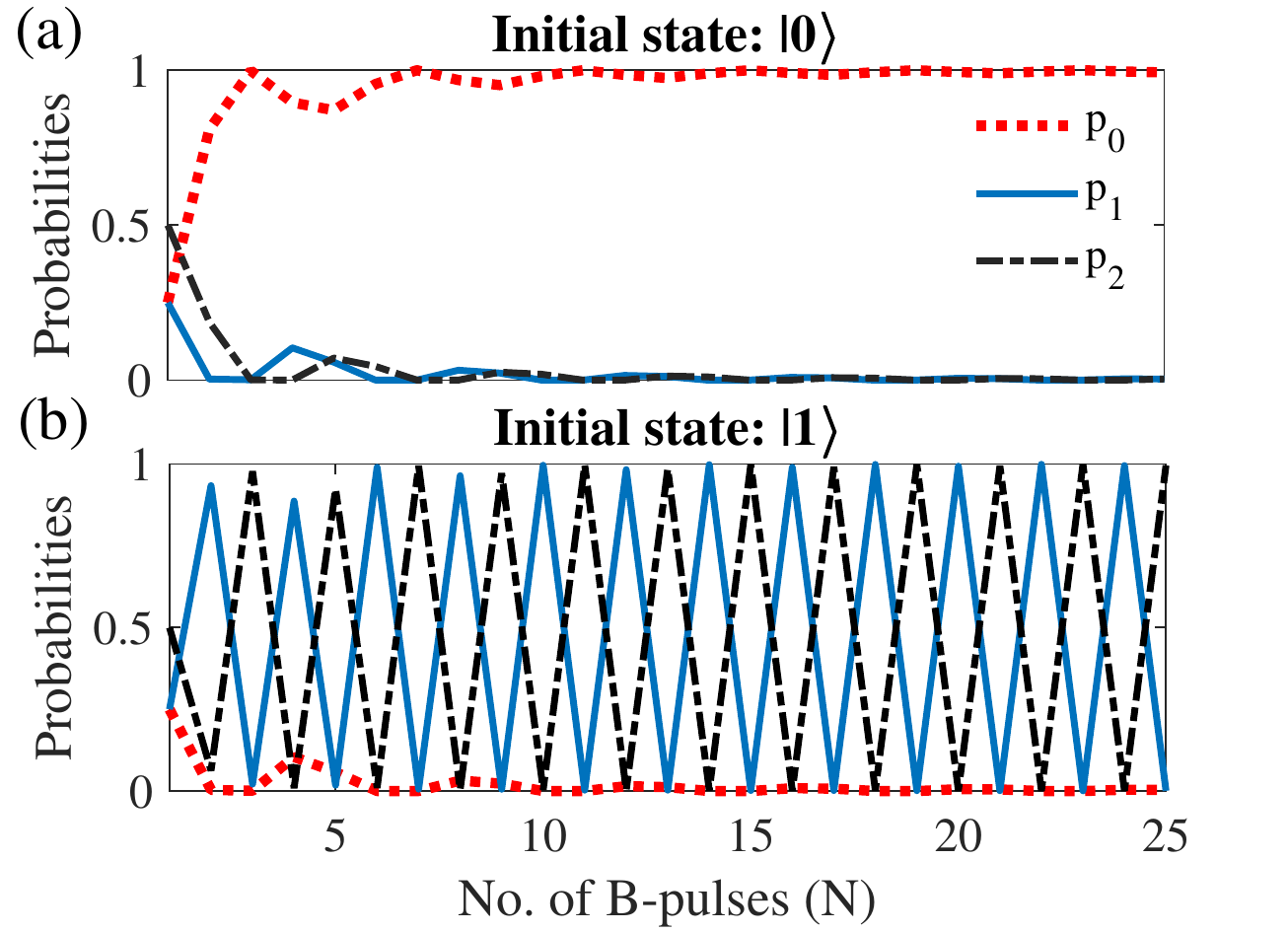}
  	\caption{Probabilities at various $N$ for $\theta = \pi$, with initial transmon state (a) $|0\rangle$ and (b) $|1\rangle$.}
  	\label{fig-initialstates}
  \end{figure}

  \subsubsection*{Quantizing the $B$-pulse in the $N>1$ case}
  
  Similarly to $N=2$, if we interrogate a single mode with $\mathtt{n}$ photons, the 3x3 matrix structure of the semiclassical case will be preserved with the replacement $|0\rangle \rightarrow |\mathtt{n}\rangle \otimes |0\rangle$, $|1\rangle \rightarrow |\mathtt{n}\rangle \otimes |1\rangle$, and $|2\rangle \rightarrow |\mathtt{n-1}\rangle \otimes |2\rangle$, as it is clear from the $N=2$ case already. Thus, all the results obtained in this paper can be applied to this situation as well.

  We can get further insights into the nature of the measurement by examining a toy-model where instead of a cavity we have a two-level system with energy levels $|\mathtt{0}\rangle$ and $|\mathtt{1}\rangle$
  that can resonantly exchange energy with the second transition of the transmon. Suppose now that we apply our protocol with a large enough $N\gg1$, with the qubit initially in a generic superposition $\alpha |\mathtt{0}\rangle + \beta |\mathtt{1}\rangle $. Based on our result so far, we would have
  \begin{equation}
  	(\alpha |\mathtt{0}\rangle + \beta |\mathtt{1}\rangle )\otimes |0\rangle \xrightarrow{N\gg 1} \alpha |\mathtt{0}\rangle \otimes |1\rangle + \beta |\mathtt{1}\rangle \otimes |0\rangle .
  \end{equation}
  This looks very similar to a CNOT gate with the qubit as the control followed by an $X$ gate on the target,
  \begin{eqnarray}
  	(\alpha |\mathtt{0}\rangle + \beta |\mathtt{1}\rangle )\otimes |0\rangle &\xrightarrow{CNOT}& \alpha |\mathtt{0}\rangle \otimes |0\rangle + \beta |\mathtt{1}\rangle \otimes |1\rangle \nonumber \\
  	&\xrightarrow{X}& \alpha |\mathtt{0}\rangle \otimes |1\rangle + \beta |\mathtt{1}\rangle \otimes |0\rangle \nonumber \\
  \end{eqnarray}
  But the similarity stops here. Indeed, the CNOT and the X gate act on the rest of the states as
  \begin{eqnarray}
  	(\alpha |\mathtt{0}\rangle + \beta |\mathtt{1}\rangle )\otimes |1\rangle &\xrightarrow{CNOT}& \alpha |\mathtt{0}\rangle \otimes |1\rangle + \beta |\mathtt{1}\rangle \otimes |0\rangle \nonumber \\
  	&\xrightarrow{X}& \alpha |\mathtt{0}\rangle \otimes |0\rangle + \beta |\mathtt{1}\rangle \otimes |1\rangle \nonumber \\
  \end{eqnarray}
  as usual. However, when using our protocol we have the following action on these states
  \begin{equation}
  	(\alpha |\mathtt{0}\rangle + \beta |\mathtt{1}\rangle )\otimes |1\rangle \xrightarrow[N {\rm odd}]{N\gg 1} -\alpha |\mathtt{0}\rangle \otimes |0\rangle -i^{N+1} \beta |\mathtt{0}\rangle \otimes |2\rangle ,
  \end{equation}
  and 
  \begin{equation}
  	(\alpha |\mathtt{0}\rangle + \beta |\mathtt{1}\rangle )\otimes |1\rangle \xrightarrow[N {\rm even}]{N\gg 1} -\alpha |\mathtt{0}\rangle \otimes |0\rangle + i^N \beta |\mathtt{1}\rangle \otimes |1\rangle ,
  \end{equation}
  The action $ |\mathtt{0}\rangle \otimes |1\rangle \rightarrow -|\mathtt{0}\rangle \otimes |0\rangle$ results immediately from $[S_{N}]^{N+1} = - i \sigma_{01}^{y} + |2\rangle \langle 2|$, implying $\mathbb{I}_{2}\otimes  [S_{N}]^{N+1} |\mathtt{0}\rangle \otimes |1\rangle  = - |\mathtt{0}\rangle \otimes |0\rangle$.
  The operations $|\mathtt{1}\rangle \otimes |1\rangle \xrightarrow[N {\rm odd}]{N\gg 1} |\mathtt{0}\rangle \otimes |2\rangle$ and  $|\mathtt{1}\rangle \otimes |1\rangle \xrightarrow[N {\rm even}]{N\gg 1} |\mathtt{1}\rangle \otimes |1\rangle$
  can be verified numerically, see for example Supplementary Fig. \ref{fig-initialstates} as well as the approximate formula for the unitary at large $N$ given in the Discussion section in the main paper.
  Again this has a straightforward physical interpretation:  after the application of an even number $N$ of pulses (that is, immediately before the $N+1$ beam-splitter), the state of the system is approximately $|\mathtt{1}\rangle \otimes |1\rangle$, that is, the qubit is excited and the transmon is in state $|1\rangle$. Since $N$ is large, after acting with the $N+1$ beam-splitter the transmon still remains approximately in the state $|1\rangle$: it can then fully absorb the excitation at the $N+1$ interaction with the qubit. This results 
  in the state $|\mathtt{0}\rangle \otimes |2\rangle$. Further on, nothing happens at the $N+2$ beam-splitter, since this acts only on the states $|0\rangle$ and $|1\rangle$. Then the $N+2$ interaction with the qubit will result in the excitation being transferred from $|2\rangle$ to the qubit. As a result, before the $N+3$ beam-splitter the state will be 
  $|\mathtt{1}\rangle \otimes |1\rangle$, which is exactly the state it entered the $N+1$ beam-splitter. The whole process then just repeats.

  This shows that our protocol is fundamentally different from the standard von Neumann measurement model, which in its simplest formulation uses a CNOT to entangle the control qubit and the target meter. Perhaps even more relevant for our problem, it is not even possible to construct a CNOT gate based only on the Hamiltonian Eq. (\ref{eq:int}), which would generate just an iSWAP type of gate. To construct a CNOT, one would need additional single-qubit gates for both the target and control \cite{nielsen-book-2002}, meaning that additional energy is exchanged, see e.g. Ref. \cite{Neumann_2010} for an explicit construction in an experiment on measuring the state of a nuclear spin.

  {\it Generalization to multiple modes.} A different scenario can be envisioned if several different modes are available, when clearly a variety of options exist on how to interrogate them. In this case, states that correspond to superpositions of these modes will be obtained when the transmon is found in the ground state, similar to what has already been observed for $N=2$. Thus, our protocol can be generalized to simultaneous detection of photons in several cavities.

  \section*{Supplementary Note 3: General results for $N \geq 1$ coherent interaction-free detection }
  
  A number of theoretical results for the case of $N\geq 1$ are presented in this section.
  
  \subsection*{General analytical results}
  
  %\textcolor{blue}{Given that the initial state is the ground state, the wavefunction after the first beam splitter is 
  	%\begin{equation}
  	%S|0\rangle = \kappa_{1}|0\rangle + \kappa_{2}|1\rangle\;,
  	%\label{eq:j=0}
  	%\end{equation}
  	%where $\kappa_{1} = \cos \frac{\pi}{2(N+1)}$ and $\kappa_{2} = \sin \frac{\pi}{2(N+1)}$.}
  
  For the coherent case, the subsequent evolution for a system of size $N$ is just $(SB)^{N}S|0\rangle$. Let us denote the wavefunction after the $j^{th}$ Ramsey segment as $|\psi\rangle_{j} = \alpha_{j}|0\rangle + \beta_{j}|1\rangle + \gamma_{j}|2\rangle$. The probability amplitudes $\alpha_{j}, \beta_{j}, \gamma_{j}$
  obey the recursion relations
  \begin{eqnarray}
  	\alpha_{j+1} &=& \cos \frac{\pi}{2(N+1)}\alpha_{j}  \label{eq:recursion_1} \\
  	& & - \sin \frac{\pi}{2(N+1)}\cos \frac{\theta_{j+1}}{2}\beta_{j}+ \sin \frac{\pi}{2(N+1)} \sin \frac{\theta_{j+1}}{2}\gamma_{j}, \nonumber \\ 
  	\beta_{j+1} &=& \sin \frac{\pi}{2(N+1)}\alpha_{j} \label{eq:recursion_2} \\
  	& & + \cos \frac{\pi}{2(N+1)}\cos \frac{\theta_{j+1}}{2}\beta_{j} - \cos \frac{\pi}{2(N+1)} \sin \frac{\theta_{j+1}}{2}\gamma_{j}, \nonumber \\
  	\gamma_{j+1} &=& \sin \frac{\theta_{j+1}}{2}\beta_{j} + \cos \frac{\theta_{j+1}}{2}\gamma_{j}\;. 
  	\label{eq:recursion_3}
  \end{eqnarray}

  In the case of identical pulses $\theta_j =\theta$, starting with the probability amplitudes Eqs. (\ref{eq:recursion_1}, \ref{eq:recursion_2}, \ref{eq:recursion_3}),  we observe that these recursion relations yield sums of even functions of $\theta$ (cosines) $\alpha_j$ and $\beta_j$, and sums of odd functions of $\theta$ (sines) $\gamma_j$.
  Specifically, the amplitudes in the coherent case can be expressed as the expansions
  \begin{eqnarray}
  	\alpha_{j} &=& \sum_{k=0}^{j} C_{j}[k]\cos \frac{k\theta}{2}, \\
  	\beta_{j} &=& \sum_{k=0}^{j} C_{j}^{'}[k]\cos \frac{k\theta}{2},\\
  	\gamma_{j} &=& \sum_{k=0}^{j} C_{j}^{''}[k]\sin \frac{k\theta}{2} .
  \end{eqnarray}

  From the recursion relations Eqs. (\ref{eq:recursion_1}, \ref{eq:recursion_2}, \ref{eq:recursion_3}),  we find the following relations among the coefficients
  \begin{eqnarray}
  	C_{j+1}[k] &=& \cos \frac{\pi}{2(N+1)} C_{j}[k]_{k=\overline{0,j}} \nonumber \\
  	&& -   \frac{1}{2} \sin \frac{\pi}{2(N+1)} \left[C^{'}_{j}[k+1] - C^{''}_{j}[k+1]\right]_{k=\overline{-1,j-1}} \nonumber \\
  	&& -   \frac{1}{2} \sin \frac{\pi}{2(N+1)} \left[C^{'}_{j}[k-1] + C^{''}_{j}[k-1]\right]_{k=\overline{1,j+1}} \nonumber \\  \\
  	C^{'}_{j+1}[k] &=& \sin \frac{\pi}{2(N+1)} C_{j}[k]_{k=\overline{0,j}}
  	\nonumber \\
  	&&+   \frac{1}{2} \cos \frac{\pi}{2(N+1)} \left[C^{'}_{j}[k+1] - C^{''}_{j}[k+1]\right]_{k=\overline{-1,j-1}} \nonumber \\
  	&& +   \frac{1}{2} \cos \frac{\pi}{2(N+1)} \left[C^{'}_{j}[k-1] + C^{''}_{j}[k-1]\right]_{k=\overline{1,j+1}} \nonumber \\ \\
  	C_{j+1}^{''}[k] &=& \frac{1}{2}\left[C^{'}_{j}[k-1] + C^{''}_{j}[k-1]\right]_{k=\overline{1,j+1}} \nonumber \\
  	&&-\frac{1}{2}\left[C^{'}_{j}[k+1] - C^{''}_{j}[k+1]\right]_{k=\overline{-1,j}} 
  \end{eqnarray}

  Here we use the notation $[ ...]_{k\overline{.. , ..}}$ to denote a restriction over the values of $k$. The final probabilities can then be easily calculated as  $p_{0} = \alpha_{N}^{2}$, 
  $p_{1} = \beta_{N}^{2}$, and  $p_{2} = \gamma_{N}^{2}$.
  The coefficients for systems of sizes $N = 1$, $N = 2$, $N = 3$, and $N = 4$ are shown in Supplementary Tables.~\ref{tab:coeffs_p0}--\ref{tab:coeffs_p2}.

  \begin{table}[H]
  	\setlength\extrarowheight{3pt}   
  	\begin{center}
  		\begin{tabular}{c||c|c|c|c|}
  			\textbf{} & \textbf{$N = 1$} & \textbf{$N = 2$} & \textbf{$N = 3$} & \textbf{$N = 4$} \\
  			\hline
  			%\hline
  			$\mathbf{C_{\textit N}[0]}$ & $0.5$ & $0.67$ & $0.75$ & $0.80 $\\[5pt]
  			\hline
  			$\mathbf{C_{\textit N}[1]}$ & $-0.5$ & $-0.43$ & $-0.36$ & $-0.31$\\[5pt]
  			\hline
  			$\mathbf{C_{\textit N}[2]}$ & $0$ & $-0.23$ & $-0.25$ & $-0.24$\\[5pt]
  			\hline
  			$\mathbf{C_{\textit N}[3]}$ & $0$ & $0$ & $-0.14$ & $-0.16$\\[5pt]
  			\hline
  			$\mathbf{C_{\textit N}[4]}$ & $0$ & $0$ & $0$ & $-0.089$\\[5pt]
  			\hline
  		\end{tabular}
  		\caption{The probability amplitude coefficients for $p_{0}$ (coherent case) up to two significant digits with sequences of length $N = 1$, $N = 2$, $N = 3$, and $N = 4$.}
  		\label{tab:coeffs_p0}	
  	\end{center}	
  \end{table}
  
  \begin{table}[H]
  	\setlength\extrarowheight{3pt}   
  	\begin{center}
  		\begin{tabular}{c||c|c|c|c|}
  			\textbf{} & \textbf{$N = 1$} & \textbf{$N = 2$} & \textbf{$N = 3$} & \textbf{$N = 4$} \\
  			\hline
  			%\hline
  			$\mathbf{C_{\textit N}^{'}[0]}$ & $0.5$ & $0.35$ & $0.27$ & $0.22$\\[5pt]
  			\hline
  			$\mathbf{C_{\textit N}^{'}[1]}$ & $0.5$ & $0.25$ & $0.17$ & $0.13$\\[5pt]
  			\hline
  			$\mathbf{C_{\textit N}^{'}[2]}$ & $0$ & $0.40$ & $0.23$ & $0.17$\\[5pt]
  			\hline
  			$\mathbf{C_{\textit N}^{'}[3]}$ & $0$ & $0$ & $0.33$& $0.21$\\[5pt]
  			\hline
  			$\mathbf{C_{\textit N}^{'}[4]}$ & $0$ & $0$ & $0$ & $0.27$\\[5pt]
  			\hline
  		\end{tabular}
  		\caption{The probability amplitude coefficients for $p_{1}$ (coherent case) up to two significant digits with sequences of length $N = 1$, $N = 2$, $N = 3$, and $N = 4$.}
  		\label{tab:coeffs_p1}
  	\end{center}	
  \end{table}

  \begin{table}[H]
  	\setlength\extrarowheight{3pt}   
  	\begin{center}
  		\begin{tabular}{c||c|c|c|c|}
  			\textbf{} & \textbf{$N = 1$} & \textbf{$N = 2$} & \textbf{$N = 3$} & \textbf{$N = 4$} \\
  			\hline
  			%\hline
  			$\mathbf{C_{\textit N}^{''}[1]}$ & $0.71$ & $0.43$ & $0.33$ & $0.27$\\[5pt]
  			\hline
  			$\mathbf{C_{\textit N}^{''}[2]}$ & $0$ & $0.47$ & $0.31$ & $0.24$\\[5pt]
  			\hline
  			$\mathbf{C_{\textit N}^{''}[3]}$ & $0$ & $0$ & $0.35$  & $0.25$\\[5pt]
  			\hline
  			$\mathbf{C_{\textit N}^{''}[4]}$ & $0$ & $0$ & $0$  & $0.29$\\[5pt]
  			\hline
  		\end{tabular}
  		\caption{The probability amplitude coefficients for $p_{2}$ (coherent case) up to two significant digits with sequences of length $N = 1$, $N = 2$, $N = 3$, and $N = 4$.}
  		\label{tab:coeffs_p2}
  	\end{center}
  \end{table}

  The recurrence relations allow us to get a deeper understanding of the process of coherent accumulation of amplitude probabilities in successive pulses. Let us consider the maximum-strength pulses $\theta_j = \pi$, for which the relations Eqs. (\ref{eq:recursion_1}, \ref{eq:recursion_2}, \ref{eq:recursion_3})  become
  \begin{eqnarray}
  	\alpha_{j+1} &=& \cos \frac{\pi}{2(N+1)}\alpha_{j}+ \sin \frac{\pi}{2(N+1)}\gamma_{j}, \label{eq:recursion_1_thetapi} \\ 
  	\beta_{j+1} &=& \sin \frac{\pi}{2(N+1)}\alpha_{j} - \cos \frac{\pi}{2(N+1)}\gamma_{j}, \label{eq:recursion_2_thetapi}\\
  	\gamma_{j+1} &=& \beta_{j}.
  	\label{eq:recursion_3_thetapi}
  \end{eqnarray}
  We notice that if the dominant probability amplitude is the one corresponding to the ground state, this relationship tends to be preserved under successive application of the sequences. Indeed, from Eq. (\ref{eq:recursion_3_thetapi}) we see that if $\beta_{j}$ is small, then $\gamma_{j+1}$ will be small as well. From Eq. (\ref{eq:recursion_2_thetapi}) we see that the relatively large probability amplitude $\alpha_j$ gets multiplied by a small number $\sin \frac{\pi}{2(N+1)}$, and the remaining part of the equation also contains the relatively small $\gamma_j$. To make this observation more precise, we note that the general form of the probability amplitudes is 
  \begin{eqnarray}
  	\alpha_{j}&=&\cos^{j+1}\frac{\pi}{2(N+1)} + \sin^{2}\frac{\pi}{2(N+1)} \mathcal{P}^{(j-2)}_{\alpha_j}\left(\cos\frac{\pi}{2(N+1)} \right), \nonumber \\ 
  	\beta_{j}&=&\sin\frac{\pi}{2(N+1)}\mathcal{P}^{(j)}_{\beta_j}\left(\cos \frac{\pi}{2(N+1)} \right), \nonumber\\ 
  	\gamma_{j}&=&\sin\frac{\pi}{2(N+1)}\mathcal{P}^{(j-1)}_{\gamma_j}\left(\cos\frac{\pi}{2(N+1)} \right).\nonumber
  \end{eqnarray}
  where $\mathcal{P}^{(j)}$ are $j$-th order  polynomials in the variable $\xi  = \cos\frac{\pi}{2(N+1)}$ satisfying 
  \begin{eqnarray}
  	\mathcal{P}^{(j-1)}_{\alpha_{j+1}}(\xi )&=& \xi   \mathcal{P}^{(j-2)}_{\alpha_j}(\xi ) + 
  	\mathcal{P}^{(j-1)}_{\gamma_j}(\xi ), \nonumber \\
  	\mathcal{P}^{(j+1)}_{\beta_{j+1}}(\xi )&=& \xi^{j+1} - \xi^{2} \mathcal{P}^{(j-2)}_{\alpha_j}(\xi ) - \xi \mathcal{P}^{(j-1)}_{\gamma_j}(\xi )  + 
  	\mathcal{P}^{(j-2)}_{\alpha_j}(\xi ), \nonumber \\
  	\mathcal{P}^{(j)}_{\gamma_{j+1}}(\xi ) &=& \mathcal{P}^{(j)}_{\beta_j}(\xi ) .\nonumber
  \end{eqnarray}
  We can see that the coefficients $\beta_j$ and $\gamma_j$ get multiplied by the small quantity $\sin \frac{\pi}{2(N+1)}$ at every iteration, therefore they tend to decrease. On the contrary, $\alpha_j$ accumulates the relatively larger quantity $\cos^{j+1}\frac{\pi}{2(N+1)}$, with $\mathrm{lim}_{N\rightarrow \infty} \cos^{N+1}\frac{\pi}{2(N+1)}=1$. Thus, at the end of the protocol, we will have $p_{0}=|\alpha_{N+1}|^{2} =
  \cos^{2(N+1)}\frac{\pi}{2(N+1)} + \sin^{2}\frac{\pi}{2(N+1)}....$. The first term equals the projective  probability, see Eq. (\ref{eq:detproj}), while the rest of terms are the result of coherent accumulation of amplitude probabilities during the sequences. We therefore expect a higher $p_0$ in the coherent case, and therefore a lower probability of absorption $p_2$. This is also calculated numerically in the next subsection.

  %%%%%%%%%
  \subsection*{Numerical results: cumulative probability of absorption}
  We have seen that the projective case of interaction-free detection completely excludes the situations where a $B$-pulse is absorbed by collapsing the wavefuction onto the state $|0\rangle$, which does not interact with the pulse.
  On the other hand, the coherent-interrogation interaction-free measurement protocol yields detection with very high probability, which is demonstrated by simulations as well as by experiments. We can introduce a figure of merit that allows us to quantify in a single number the probabilities of $B$-pulse absorption at different sequences.
  We can quantify this concisely by keeping track of the probability of absorption instances with $\theta=\pi$ at each sequence $j \in [1,N]$. For a given $N$ we introduce $\mathbb{C}=\sum_{j=1}^N p_2 (j)$, which essentially quantifies cumulatively the unfavorable absorption events.
  In Supplementary Fig.~\ref{fig:fig7}, the black curve corresponds to the cumulative probability with which photons can get absorbed in a projective measurement protocol and the blue curve corresponds to the total probability obtained by adding the state-$|2\rangle$ probabilities at the end of each $B$-pulse in the coherent measurement protocol. It is clearly seen that the coherent measurement protocol has less cumulative net probability of $B$-pulse absorption. 
  %This counter-intuitive behavior is due to same phase of all the $B$-pulses, which assigns phases of zero or $\pi$ to the cumulative coefficients of level $|2\rangle$ and hence lead to addition or subtraction of the cumulative coefficients in the quantum state of a three-level system.
  \begin{figure}[ht]
  	\centering
  	\includegraphics[width=8cm,keepaspectratio=true]{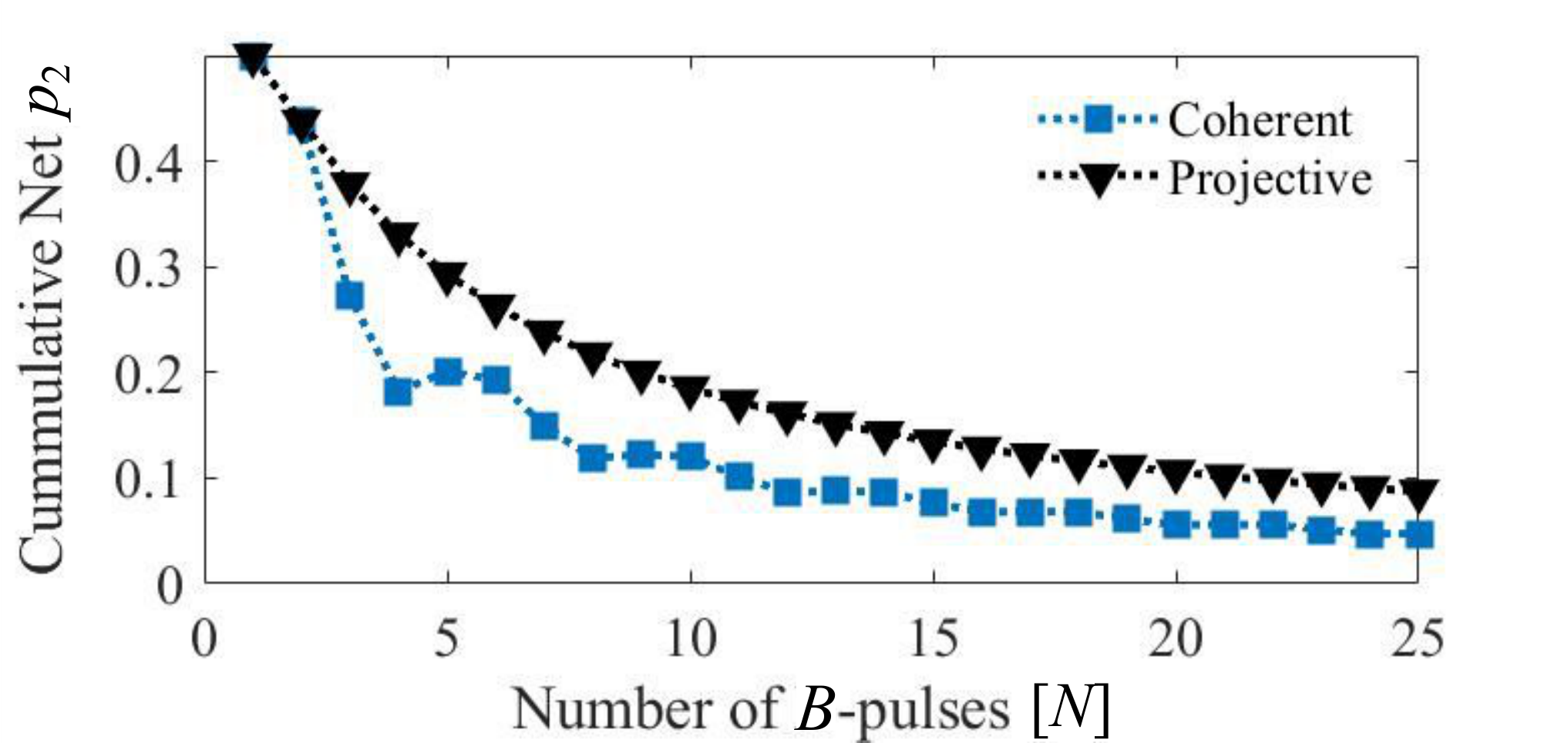}
  	%\includegraphics[width=8cm,keepaspectratio=true]{cost4.jpg}
  	% MZI_revision_fig1.jpg: 510x511 px, 72dpi, 17.99x18.03 cm, bb=0 0 510 511
  	\caption{Cumulative probability $\mathbb{C}$ of $B$-pulse absorption with $\theta=\pi$ versus $N$. The blue dotted curve with square markers corresponds to the case of the coherent interaction-free measurement protocol and the black dotted curve with triangular markers results from the projective measurement protocol.}
  	\label{fig:fig7}
  \end{figure}

  \subsection*{Identical and random pulses}
  %%%%%%%%%%%%%%%%%%
  \begin{figure}[h]
  	\centering
  	\includegraphics[width=0.9\linewidth]{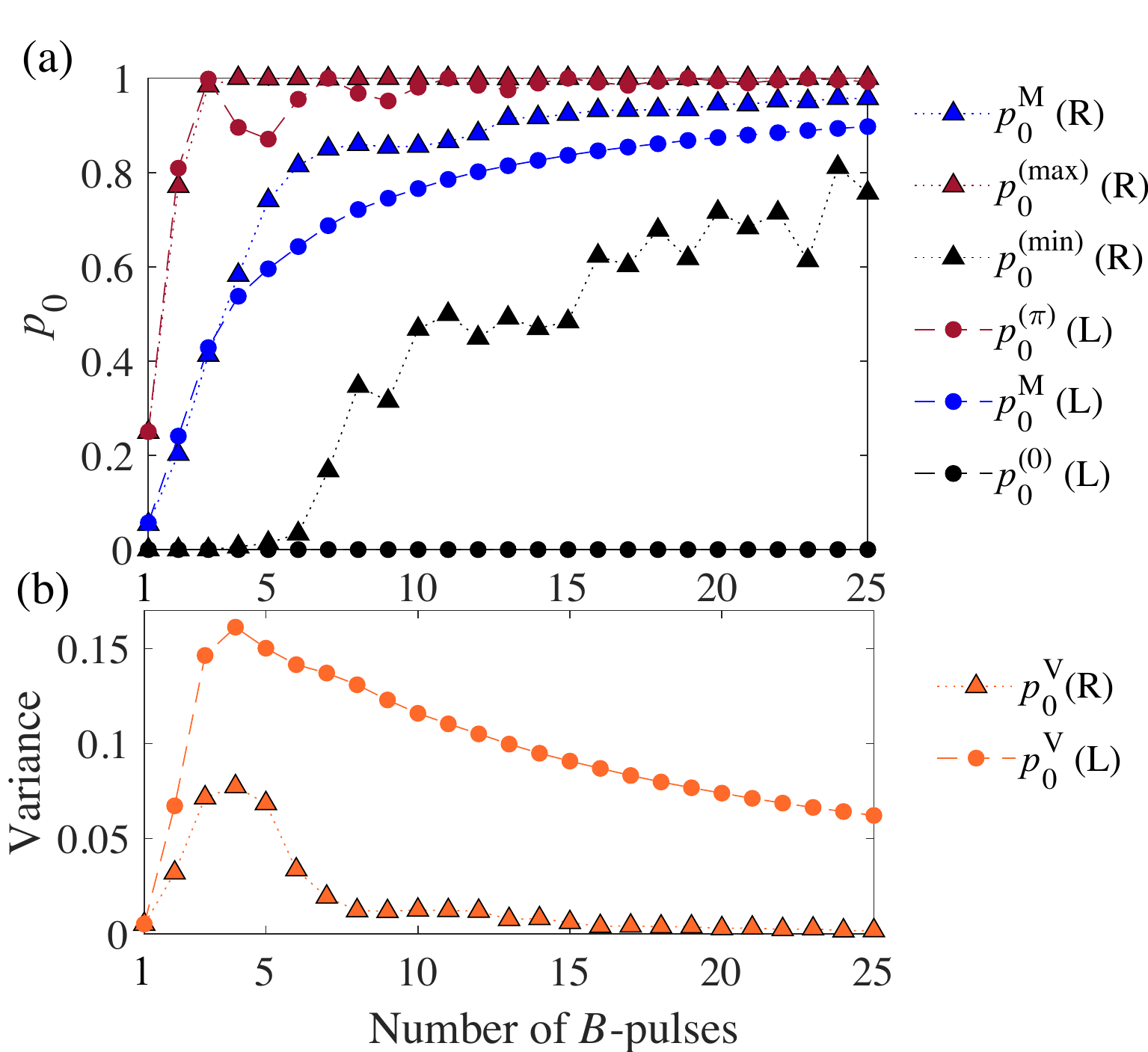}
  	\caption{(a) Mean and extreme values of $p_0$ are plotted for 
  		$N \in [1,25]$ for the cases of identical $B$-pulses 
  		($\theta_i = \theta_j = \theta \, i,j \in [1,N]$) as $\theta$ varies 
  		linearly from $0$ to $\pi$ for each $N$ with circular markers. Triangular 
  		markers present the corresponding results when $\theta_i \neq \theta_j$ and 
  		each $\theta_i$ is chosen arbitrarily from $[0,\pi]$ independent of any $\theta_j$. $p_0^{\rm M}$(L), $p_0^{(0)}$(L), and $p_0^{(\pi)}$(L)
  		are mean, minimum and maximum values respectively of the $p_0$ distribution for the case of identical $B$-pulses with linearly vaying strengths.
  		$p_0^{\rm M}$(R), $p_0^{\rm (min)}$(R), and $p_0^{\rm (max)}$(R)
  		are mean, worst and best values respectively from the $p_0$ distribution for the case of random $B$-pulse strengths. (b) Curves showing the corresponding variance of the $p_0$ distributions with $N$.}
  	\label{fig:idealcase_nbombs}
  \end{figure}
  %%%%%%%%%%%%%%%%%%%%%%%%%
  Let us have a closer look at the simulation in the case of $N$ $B$-pulses with 
  equal and unequal (random) pulse strengths. In Supplementary Fig. \ref{fig:idealcase_nbombs},
  circular markers present the case of $N \in [1,25]$ $B$-pulses with equal 
  strengths and triangular markers correspond to randomly chosen $B$-pulse strengths. In this case, the strengths $\theta$ of each $B$-pulse increase linearly from $0$ to $\pi$ in $400$ steps and the resulting distributions of 
  the ground state 
  probability $p_0$ is obtained. As expected, in Supplementary Fig. \ref{fig:idealcase_nbombs}(a) the black circles connected by the dashed black line representing the case of no $B$-pulses yields $p_0=0$, while the dashed line with 
  red circular markers corresponds to $\theta=\pi$, which has a tendency to stay closer to 1. We note that $\theta=\pi$ may not correspond to the maximum value of $p_0$, especially for smaller values of $N$. In fact the maximum $p_0$ in both cases (equal and unequal $B$-pulse strengths) coincide with each other and is represented by red triangles.
  The average value of ground state probability $p_0^{\rm M}$(L) in the case of 
  linearly varying $\theta$ gradually increases from $0.057$ for $N=1$ to $0.897$
  for $N=25$ as shown with the blue dashed curve with circles. Interestingly, 
  the situation with randomly chosen $\theta \in [0,\pi]$ 
  (400 samples for each $N$), gives rise to 
  higher average values ($p_0^{\rm M}$(R)) as shown with the blue dotted curve with 
  triangular markers. Black and red dotted curves with triangular markers 
  result from the worst and best combinations of random $B$-pulses. It is noteworthy
  that even the worst choice of random $B$-pulses have a good chance of being detected. 
  While the ignorance about the $B$-pulse strengths appear to benefit in this case, 
  results from randomly chosen $B$-pulse strengths also depend upon the sample size (here the sample size is 400). Further, variance of the $p_0$
  distributions for each $N$ is shown in Supplementary Fig.~\ref{fig:idealcase_nbombs}(b),
  where circular markers correspond to the case of equal $B$-pulse strengths and 
  triangular markers correspond to the case of random $B$-pulse strengths. Much 
  lower values of variance are obtained in the case of arbitrarily chosen $B$-pulse 
  strengths.
  %%%%%%%%%%%%%%%%

  %%%%%%%%%%%%%%
  \section*{Supplementary Note 4: Discussion: ignorance is bliss}

  The previous numerical simulations demonstrate that the coherent case is more efficient than the standard projective (quantum Zeno effect) case. This is a non-intuitive result, because negative measurements, while not producing any macroscopic event (detector click, etc.) still provide more information. A famous example outside quantum physics is the Monty Hall problem. 
  
  However, the strategy of extracting ``classical'' information is not necessarily advantageous, as the case of coherent interaction-free detection realized in this paper demonstrates. To give a qualitative justification of why it is so, let us consider the state $\sqrt{1-x^2-y^2}|0\rangle + x |1\rangle + y |2\rangle $ at the input of a Ramsey segment containing the pulse $B(\pi )$. After going through the interferometer the probability $p_{2}$ of the state $|2\rangle$ is $\left[\sqrt{1-x^2-y^2}\sin (\pi/2(N+1)) + x\cos(\pi/2(N+1))\right]^2$. Let's examine now the projective scenario. In this case, the input state should not contain any component on the state $|2\rangle$, since in this protocol the state is always projected on the $\{|0\rangle , |1\rangle \}$ subspace. Considering $\sqrt{1-x^2}|0\rangle + x |1\rangle$ as the input state, we find that the probability of detection (``explosion'') is  $\left[\sqrt{1-x^2}\sin (\pi/2(N+1)) + x\cos(\pi/2(N+1))\right]^2$, clearly larger than in the coherent case.

  \section*{Supplementary Note 5: Experimental errors due to pulse imperfections}
  We present here an analysis of errors due to the imperfect generation of pulses in our setup. These imperfections are: IQ mixer saturation, finite sampling rates, detunings with respect to the corresponding transition frequencies, etc. For example, IQ mixer saturation effects start to be observable in our setup for values $\theta > 3\pi$ (approximately); at the highest power, a pulse with amplitude $\Omega_0(4\pi)$ in fact implements a unitary with $\theta=3.9\pi$. These imperfections are embedded in our simulations.

  To characterize these errors we obtain the explicit form 
  of the unitary evolution generated by the drive Hamiltonian 
  in Eq.~3 (from the main text) and compare it with the ideal beam-splitter 
  unitaries $S_1$ (Eq.~\ref{Eq-S1}), $S_2$ (Eq.~\ref{Eq-S2}),
  and the $B$-pulses $B(\theta)$ (Eq.~\ref{Eq-B}).
  The Hamiltonian in Eq.~3 (from the main text) generates a unitary
  evolution, and the corresponding dynamics can be determined 
  by solving
  \begin{equation}
  	i \hbar \frac{\partial}{\partial t} \vert \psi(t) \rangle = H(t) \vert \psi(t) \rangle, \label{eq:SchroedingerEq}
  \end{equation}
  where $\vert \psi(t) \rangle$ is the state of the system at 
  an arbitrary time $t$ and $\Omega_{01}(t)$ ($\Omega_{12}(t)$)
  are the drives from Eq. 1 (main text). 
  Following Eq. (\ref{eq:SchroedingerEq}), we obtain the dynamics 
  of our three-level system initialized in the computational 
  basis states $\vert 0 \rangle$, $\vert 1 \rangle$, and $\vert 2 \rangle$. Each of these states undergo the same evolution for the same time, resulting in states $\vert \psi_0(t) \rangle$, $\vert \psi_1(t) \rangle$, and $\vert \psi_2(t) \rangle$
  respectively. The explicit form of the corresponding unitary operator 
  at arbitrary time $t$ can thus be written as
  \begin{equation}
  	U_{\rm sim}(t)= \vert \psi_0(t) \rangle \langle 0 \vert
  	+ \vert \psi_1(t) \rangle \langle 1 \vert
  	+ \vert \psi_2(t) \rangle \langle 2 \vert.
  \end{equation}
  This unitary operator is obtained numerically for given experimental parameters. Numerical integration is performed using the fourth-order Runge-Kutta method with a step size of 
  (AWG sampling rate)$^{-1}$. Thus, the super-Gaussian pulse envelope of $56$ ns duration is discretized by the AWG sampling rate of $1$ GS/s.
  Deviation of $U_{\rm sim}$ from 
  the ideally expected $U(t)$ ($=S_1$ or $S_2$ or $B(\theta)$)
  is calculated using the 2-norm of the difference between the operators, given by $\xi = ||U(t)-U_{\rm sim}(t)||_2$. Note that the maximum value of $\xi$ is 2. 
  Thus we can quantitatively assess each
  individual unitary operation implemented in the experiments.
  We have also analyzed the results of  single-qutrit quantum process tomography (QPT)~\cite{nielsen-book-2002, qpt_2021} 
  and obtained the precision
  of the overall pulse sequence ($S_1 B(\theta) S_1$). 
  Process matrices ($\chi_{\rm sim}$) resulting from the simulation (without decoherence) including 
  pulse errors are compared with that of the ideal process 
  matrix ($\chi_{\rm ideal}$) using the fidelity measure
  $\mathcal{F}= [\textrm{Tr}\sqrt{\sqrt{\chi_{\rm ideal}} \chi_{\rm sim} \sqrt{\chi_{\rm ideal}}}]^2$. 
  
  Overall, this analysis results in the following bounds for the errors. For $N=1$
  the beam-splitter unitary $S_1$ has $\xi=0.01$ and the average $\xi$ for $B(\theta)$ is $0.06$. From tomography, $\mathcal{F}=0.98$, averaged over $\theta \in [0, 4\pi]$. For $N=2$, we have $\xi=0.07$ for $S_2$ and average $\xi=0.21$ for $B(\theta)$. Further, evaluating the whole process ($S_2 B(\theta) S_2 B(\theta) S_2$) via QPT, we 
  obtain an average fidelity $\mathcal{F}=0.91$. Results from the simulations with these 
  pulse errors alongside with decoherence match the experimental datasets quite well as shown in Figs. (2,3) of the main text and Supplemetary Fig. (\ref{fig:012}).

  %%%%%%%%%%%%%%
  \section*{Supplementary Note 6: Representation on the Majorana sphere}

  Geometrical representations are useful for understanding quantum operations. 
  Here we adopt the Majorana representation to visualize geometrically
  the single-qutrit dynamics during our protocol. We simulate the single-qutrit dynamics on the Majorana sphere for the case of multiple consecutive MZI setups.

  In the Majorana geometrical representation, a particle with spin $j$ is represented by $2j$ points (known as the Majorana stars) on a unit sphere (known as the Majorana sphere). Consider an arbitrary state of a spin $j$ particle in the $|jm\rangle$ basis, 
  \begin{equation}
  	|\Psi \rangle = \sum_{m=-j}^{j} c_{m} |jm\rangle, 
  \end{equation}
  where $c_m$ are the complex coefficients. The corresponding Majorana polynomial
  of degree $2j$ is constructed as $P_{|\Psi\rangle} (\zeta ) = a_0 \zeta^{2j} + a_1 \zeta^{2j-1} +\dots + a_{2j}$, with
  \begin{equation}
  	a_r=(-1)^r \frac{c_{j-r}}{\sqrt{r!}\sqrt{(2j-r)!}}.
  \end{equation}
  $P_{|\Psi\rangle} (\zeta )=0$ has $2j$ roots, which can be plotted 
  in the  $xOy$ plane. The inverse stereographic
  projections of each of these points with respect to the South Pole of the 
  unit sphere give rise to Majorana stars ($\mathcal{S}_i$, $i \in [1,2j]$). Thus, the Majorana representation of a qutrit ($j=1$) consists of two Majorana stars.
  For $j=1$ we have $m=-1, 0, 1$, and the qutrit basis $|0\rangle$, $|1\rangle$, $|2\rangle$
  may be identified as $|1m\rangle \equiv |1-m\rangle$.
  For the state $|0\rangle$ both Majorana stars lie on the North Pole,  
  $|2\rangle$ has both Majorana stars lying on the South Pole,
  while $|1\rangle$ is represented by one Majorana star on the North Pole and another one on the South Pole. 
  
  %%%%%%%%%%%%%%%%%%%%%%%%%%%%
  \begin{figure}
  	\centering
  	\includegraphics[scale=1,keepaspectratio=true]{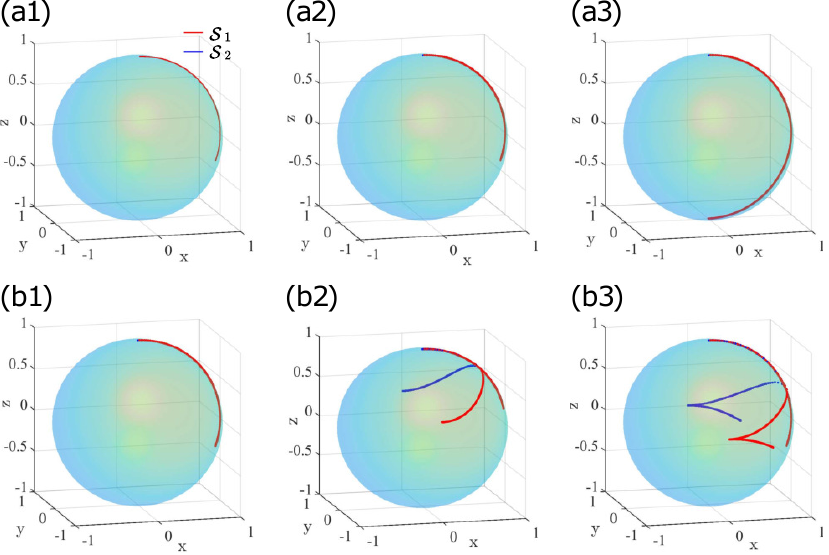}
  	\caption{Majorana trajectories at each step in the single Ramsey protocol without a pulse (a1)-(a2)-(a3) and with a pulse (b1)-(b2)-(b3) using a three-level quantum system. 
  	} \label{Fig-Maj1}
  \end{figure}

  \begin{figure}
  	\centering
  	\includegraphics[scale=1,keepaspectratio=true]{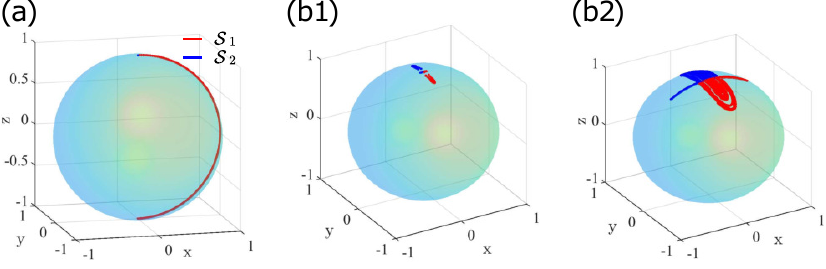}
  	\caption{Majorana trajectories resulting from the emulation of $N$ consecutive MZI setups with Ramsey angles $\pi/(N+1)$.
  		The subplots correspond to:
  		(a) no  pulse, (b1) $N=100$ with 100 equal-strength pulses ($\theta_j=\pi$), and (b2) $N=100$ with 100 pulses ($\theta_j=$random).} \label{Fig-Maj2}
  \end{figure}
  %%%%%%%%%%%%%%%%%%%%%%%%%%%%
  
  First, we consider a single Ramsey setup. We
  initialize a qutrit in the state $|0\rangle$ and simulate the pulse 
  sequence for $\theta_1=0,\pi$. The corresponding quantum state dynamics is calculated and plotted as a dynamics of Majorana stars in Supplementary Fig.~\ref{Fig-Maj1}. Supplementary Figs.~\ref{Fig-Maj1}(a1),(a2),(a3) present 
  the trajectory under the first beam splitter, during the evolution in the absence of a pulse, and respectively under the second
  beam splitter. The Majorana stars
  $\mathcal{S}_1(x_1,y_1,z_1)$ and $\mathcal{S}_2(x_2,y_2,z_2)$ are shown in
  red and blue colors, where $x_i,y_i,z_i$ are Cartesian coordinates. 
  To begin with, both stars lie at the North Pole,
  corresponding to the state $\vert 0 \rangle$. Under the effect of the first beam splitter, the Majorana star $\mathcal{S}_1$ moves in the plane $y=0$, while $\mathcal{S}_2$
  stays at the North Pole such that the qutrit attains the state: $(|0\rangle+|1\rangle)/\sqrt{2}+|2\rangle$, see Supplementary Fig.~\ref{Fig-Maj1}(a1). Further, since there is no pulse in this case, no change is observed in Supplementary Fig.~\ref{Fig-Maj1}(a2). Finally, the second beam splitter brings $S_1$ to the South Pole of the sphere, see Supplementary Fig.~\ref{Fig-Maj1}(a3), thus representing the state $|1\rangle$ (one star at the north pole, one star at the south pole). The corresponding trajectories for $\theta_1=\pi$ 
  are shown in Supplementary Fig.~\ref{Fig-Maj1}(b1),(b2),(b3). These two cases
  with $\theta_1=0,\pi$ are clearly distinct as observed from Supplementary Figs.~\ref{Fig-Maj1}(a2) and (b2). Supplementary Fig.~\ref{Fig-Maj1}(b2) shows a non-trivial trajectory, in which 
  $\mathcal{S}_1$ partially retraces its path very quickly and in the meantime $\mathcal{S}_2$ 
  moves along the previous trajectory of $\mathcal{S}_1$ such that both of these 
  Majorana stars meet somewhere in the middle of the trajectory and then 
  start moving symmetrically in different directions. This step 
  corresponds to the generation of the coherence between states $|0\rangle$ 
  and $|2\rangle$. Finally, the last step implements the same beam splitter again,
  leading to the state $(|0\rangle+|1\rangle)/2+|2\rangle/\sqrt{2}$ with Majorana stars $\mathcal{S}_1(0.586,0.792,-0.172)$ and $\mathcal{S}_2(0.586,-0.792,-0.172)$. Clearly,
  the case $\theta \neq 0$  and $\theta=\pi$ are distinguishable by 
  the different star constellations.

  Next, we proceed with this geometrical representation and observe the single qutrit dynamics with multiple pulses.
  For $N=2$, with $\theta_1=\theta_2=\pi$, we find that the coordinates 
  of the final-state Majorana stars are $(0.062,\pm 0.935,0.350)$.
  For $N\geq2$, both the Majorana stars end up in the northern hemisphere. 
  %\textcolor{blue}{}
  Supplementary Figs.~\ref{Fig-Maj2}(a-b1-b2) present
  the final states
  obtained in the case of no pulse, 100 pulses with equal $\theta_j=\pi$, and 100 pulses with randomly chosen $\theta_j$s respectively. As discussed earlier, the case of no pulse corresponds to the Majorana stars
  $\mathcal{S}_1(0,0,-1)$ and $\mathcal{S}_2(0,0,1)$.

  The final state of the single-qutrit emulating 100 consecutive Ramsey setups with 100 pulses is confined to the region around the North Pole, see Supplementary Figs.~\ref{Fig-Maj2}(b1-b2). Thus a completely contrasting configuration
  of the Majorana stars is observed for the case of no pulse versus the case 
  with many pulses (here 100). In Supplementary Fig.~\ref{Fig-Maj2}(b1), where all $B$-pulses are of strength $\theta=\pi$, final state has both Majorana stars lying very close to the North pole. An interesting situation is seen in Supplementary Fig.~\ref{Fig-Maj2}(b2) wherein even a bad choice of arbitrary strengths of the $B$-pulses also correspond to a Majorana trajectory that is found to stay close to the North pole. An example showing the average Majorana trajectory with arbitrary $B$-pulse strengths for $N=25$ is shown in Fig.~7 of the main text. Thus it is clear from the Majorana geometrical representation that for the case of large number of pulses, the
  probability of interaction-free detection is quite high and that the $B$-pulse strength does not matter anymore.

  To conclude, we obtained the signature of coherent interaction free detection on the 
  Majorana sphere. The results obtained from the single qutrit dynamics
  on the Majorana sphere is in complete agreement with the theoretical 
  expectations and simulations.


\end{document}
%
% ****** End of file apssamp.tex ******